\documentclass[a4paper,11pt]{article}

\usepackage{amsmath,amsfonts,amssymb,caption,graphicx}
\usepackage{ushort}
\usepackage{makeidx}
\usepackage{float}
\usepackage{accents}
\usepackage{color}
\usepackage{xcolor}
\usepackage{framed}
\usepackage{versions}
\usepackage{emptypage}
\usepackage{wrapfig}
\usepackage[T1]{fontenc}
\usepackage[utf8]{inputenc}
\usepackage{titlesec}
\usepackage{fancyhdr}
\usepackage{extramarks}
\usepackage[bookmarks]{hyperref}
\usepackage{hyperref}
\usepackage{bbm}
\usepackage{enumitem}
\usepackage[page]{appendix}
\usepackage{subcaption}
\usepackage{mathrsfs}
\usepackage[page]{appendix}

\usepackage{pgfplots}
\pgfplotsset{compat=1.18}
\usepackage{amsmath}
\usepackage{tikz}
\usetikzlibrary{patterns}
\usepgfplotslibrary{fillbetween}

\usepackage{ragged2e}

\hypersetup{
    colorlinks=true,   	
    linkcolor=red,      
    citecolor = [rgb]{0 0.7 0},   	
    filecolor=magenta, 	
    urlcolor=blue
}

\newcommand{\RL}{{\mathbb R}}

\newcommand{\IND}{{\mathbb I}}
\newcommand{\BBP}{{\mathbb P}}

\newcommand{\BBE}{{\mathbb E}}

\def\ba{\begin{align}}
\def\ea{\end{align}}
\def\ban{\begin{align*}}
\def\ean{\end{align*}}

\def\be{\begin{eqnarray}}
\def\ee{\end{eqnarray}}
\def\ben{\begin{eqnarray*}}
\def\een{\end{eqnarray*}}

\def\bqq{\begin{equation}}
\def\eqq{\end{equation}}
\def\bqqn{\begin{equation*}}
\def\eqqn{\end{equation*}}

\def\sq{$\Box$}

\def\qed{\ifmmode\sq\else{\unskip\nobreak\hfil
\penalty50\hskip1em\null\nobreak\hfil\sq
\parfillskip=0pt\finalhyphendemerits=0\endgraf}\fi\par\medbreak}

\newsavebox{\junk}
\savebox{\junk}[1.6mm]{\hbox{$|\!|\!|$}}

\def\limsup{\mathop{\rm lim\ sup}}
\def\liminf{\mathop{\rm lim\ inf}}

\def\til={{\widetilde =}}

\def\clA{{\cal A}}
\def\clB{{\cal B}}
\def\clC{{\cal C}}

\def\clH{{\cal H}}

\def\clW{{\cal W}}

 \def\eq#1/{(\ref{#1})}

\newtheorem{theorem}{Theorem}[section]
\newtheorem{corollary}[theorem]{Corollary}
\newtheorem{proposition}[theorem]{Proposition}
\newtheorem{lemma}[theorem]{Lemma}

\def\eq#1/{(\ref{e:#1})}

\def\bdes{\begin{description}}
\def\edes{\end{description}}

\def\notes#1{}

\definecolor{mag}{rgb}{0.7,0,0.3}
\definecolor{dgreen}{rgb}{0.1,0.5,0.1}
\definecolor{dred}{rgb}{.8,0,0}
\definecolor{gray}{rgb}{.8,.8,.8}
\definecolor{brown}{rgb}{0.6451,0.3706,0.1745}

\DeclareUnicodeCharacter{03B1}{$\alpha$}

\title{The Sample Complexity of Channel Coding}

\author{
Terence Viaud
\and 
Ioannis Kontoyiannis
}

\date{\today}

\begin{document}

\maketitle

\footnotetext{The authors are with the
Statistical Laboratory, Centre for Mathematical Sciences, 
University of Cambridge, Wilberforce Road, Cambridge CB3 0WB, 
UK. Email: 
\texttt{tv329@cam.ac.uk},
\texttt{yiannis@maths.cam.ac.uk}.\\
This work was supported in part
by the EPSRC-funded INFORMED-AI project EP/Y028732/1.}

\begin{abstract}
We introduce a new framework for the study
and evaluation
of the fundamental limits of channel coding,
emphasizing genuinely nonasymptotic results. 
The {\em channel coding sample complexity} 
$n^*(\clW,\epsilon)$ of a channel $\clW$
is defined as the smallest 
blocklength at which it is possible to transmit 
information through that channel
with error probability below~$\epsilon$
and at a rate at least as high as that
given by an appropriate constraint also 
in terms of~$\epsilon$.
This formulation follows analogous developments
in statistics, computer science and, more recently, 
in lossless source coding. 
For any discrete memoryless channel we derive
nonasymptotic bounds on the sample complexity,
with explicit constants.
These bounds show that the sample complexity
is characterized not by the usual channel capacity,
but by the channel's Augustin--Csisz\'{a}r capacity 
of order~$1/2$, which coincides with the cutoff rate 
as well as with the 
channel's Sibson capacity of order~$1/2$.
The proof is partly based on 
a new channel-coding one-shot converse 
in terms of hypothesis testing. 
We also examine extensions of the sample complexity
framework to general target rates.
Moreover,
we derive corresponding results 
for the sample complexity of the 
additive white Gaussian noise channel 
with a quadratic power constraint.
Their proof relies on new, explicit upper bounds 
on the best achievable error probability,
which may be of independent interest.
\end{abstract}


\newpage

\section{Introduction}

\subsection{Channels and codes}
\label{s:channelcodes}

A {\em channel} $W=(W_{Y|X})$ {\em from $A$ to $B$} 
is a family of regular conditional 
distributions for $Y$ given $X$, for two
random variables $X,Y$ taking values in Polish spaces, 
the {\em input alphabet} $A$ and 
the {\em output alphabet} $B$, respectively.

A {\em channel code} $(C,d)$ for a channel $W$ consists of
$(i)$~a {\em codebook} $C\subset A$ consisting
	of $M \geq 1$ distinct {\em codewords}
	$x(1),\ldots,x(M)$, 
	with each 
	codeword $x(j)$ indexed by {\em message} $j\in
	\{1,2, \ldots, M\}$;
	and $(ii)$~a {\em decoder} $d : B \rightarrow \{1,2\ldots,M\}$, 
	which is a possibly randomized, measurable map from the output
	alphabet $B$ to the set of messages $\{1,2,\ldots,M\}$.

Given a code $(C,d)$  and a message $j$,
the probability of error when 
transmitting $j$ is 
$P_e(j) 
=  \BBP(d(Y)\neq j|X=x(j))$, and
the overall probability of error of the code is measured
by either the {\em maximum probability of error}, 
    $$P_{e,\max} := \max_{1\leq j\leq M} P_e(j),$$
where $M=|C|$ is the size of the code, 
or by the {\em average probability of error}, 
    $$P_{e,{\rm avg}} := \frac{1}{M} \sum_{j=1}^M P_e(j).$$

In the typical case where communication takes place
in repeated transmissions,
a {\em channel} $\clW=\{(W_{Y^n|X^n})\;;\;n\geq 1\}$ 
from $A$ to $B$ is sequence of 
families of conditional distributions $(W_{Y^n|X^n})$ 
with input alphabet $A^n$ and output alphabet $B^n$.
When each $(W_{Y^n|X^n})$ is the product of a 
particular channel $W=(W_{Y|X})$ from $A$ to $B$, 
we say that the channel $\clW=\{W^n\}$ is
{\em memoryless}. A {\em discrete memoryless channel} (DMC) $\clW=\{W^n\}$
is a memoryless channel with discrete input and output alphabets.
The {\em rate} of a code $(C_n,d_n)$ of blocklength $n$
with a codebook $C_n$ consisting of 
$M_n=|C_n|$ codewords $x^n(j)=(x_1(j),\ldots,x_n(j))\in A^n$,
$1\leq j\leq M_n$ for a channel $(W_{Y^n|X^n})$,
is 
$$R_n=\frac{1}{n}\log M_n\quad\mbox{bits per transmission},$$
and its associated error probabilities are defined as before and
denoted by
$P_{e,{\rm max}}^{(n)}$
and
$P_{e,{\rm avg}}^{(n)}$.

The best achievable performance of channel codes is naturally 
examined in terms of the trade-off between high rates
and low error probabilities.
Formally, for any rate $R\geq 0$ and blocklength~$n$, the best
achievable average error probability is
\be
\epsilon^*(n,R) :=
\inf\left\{
P_{e,{\rm avg}}^{(n)}\,:\,
{\rm codes}\;(C_n,d_n)\;{\rm with}\;\log |C_n|\geq nR
\right\}.
\label{eq:epsilonstar}
\ee
Similarly, the 
maximal code size achievable with 
average error probability 
no greater than $\epsilon\in(0,1)$ is
$$M^*(n,\epsilon) := \sup \left\{ M \geq 1 : 
\epsilon^*\Big(n,\frac{1}{n}\log M\Big)
\leq \epsilon
\right\}.$$

\subsection{Fundamental limits and expansions}
\label{s:existingapproaches}

Suppose $\clW=\{W^n\}$ is a memoryless channel.
In the vast majority of cases
the exact evaluation of the 
fundamental limits of $M^*(n,\epsilon)$ and $\epsilon^*(n,R)$
is extremely difficult or impossible. Therefore,
most prior theoretical work 
has focused on large-blocklength asymptotic approximations.
The first-order behaviour of $M^*(n,\epsilon)$ was already
determined by Shannon~\cite{shannon:48,gallager:book}
who showed that, 
$$\log M^*(n,\epsilon)=n C(\clW)+o(n).$$
This holds
for any $\epsilon\in(0,1)$,
as the blocklength $n\to\infty$, where $C(\clW)$ is
the channel's {\em capacity}.
For a class of channels including DMCs
and Gaussian channels, this was refined
by Polyanskiy, Poor and Verd\'{u}~\cite{PPV:10} who,
following early work by Strassen~\cite{strassen:64b},
showed that
\be 
\log M^*(n,\epsilon) = nC(\clW) - \sqrt{nV(\clW)} 
Q^{-1}(\epsilon) + O(\log n),
\label{eq:expansion1}
\ee
where $V(\clW)$ is the channel {\em dispersion}
and $Q(z) = 1 - \Phi(z)$, $z \in \mathbb{R}$, is the 
standard Gaussian tail function. 

A corresponding program has been carried out for the asymptotics
of the optimal error probability $\epsilon^*(n,R)$.
After early results by Rice~\cite{rice:50}, Feinstein~\cite{feinstein:54}, 
Elias~\cite{elias:55} and Shannon~\cite{shannon:57b},
who provided conditions under which $\epsilon^*(n,R)$ 
decays exponentially 
with $n$ at all rates $0<R<C(\clW)$, i.e.,
$$\epsilon^*(n,R)=2^{-n E^*(R)+o(n)},\qquad E^*(R)>0,$$
a lot of work has been devoted to 
identifying the optimal exponent
given by the {\em reliability function} $E^*(R)$,
and the polynomial prefactors of $\epsilon^*(n,R)$.
For example, the best exponent 
$E_r(R):=\sup_{P,\rho\in[0,1]}[E_0(\rho,P)-\rho R]$ 
in the random coding upper bound 
(cf.~Proposition~\ref{prop:rcu}) and 
the exponent in the sphere packing lower bound~\cite{gallager:book} 
coincide for rates above the {\em critical rate} $R_c$,
so $E^*(R)=E_r(R)$ for $R_c<R<C(\clW)$.

For more recent relevant developments
see, e.g.,~\cite{scarlett:13,tan:15,altug:21}
and the references therein.

\subsection{Sample complexity}
\label{s:intronstar}

In this work,
we introduce a different, genuinely nonasymptotic approach to quantifying
the fundamental performance limits of channel coding. This approach is 
partly motivated by parallel developments in the statistics and 
computer science literature, and more recently in lossless
data compression; 
see~\cite{viaud:ISIT26,viaud:26ITT} for an extensive 
historical overview.

Broadly speaking
the goal of channel coding is to select a code 
with a large codebook and small probability of error.
So, for any given channel $\clW$, it makes sense to
define 
the sample complexity $n^*$
as the shortest blocklength at which such codes exist. 
To be precise, for any channel $\clW=\{(W_{Y^n|X^n})\}$ 
and any $\epsilon\in(0,1)$, 
we define {\em the channel coding sample complexity} 
$n^*_{\rm avg}(\clW,\epsilon)$ of $\clW$ with respect to 
the average error probability as
\be
n^*_{\rm avg}(\mathcal{W},\epsilon) 
:= \inf \left\{ n \geq 1 : \inf_{(C_n,d_n)} 
\max \Big\{ P^{(n)}_{e,{\rm avg}},\frac{1}{|C_n|} \Big\} 
\leq \epsilon \right\},
\label{def:scavg}
\ee
where the infimum is taken over all codes
$(C_n,d_n)$ of blocklength $n$.
Similarly, 
{\em the channel coding sample complexity} 
$n^*_{\rm max}(\mathcal{W},\epsilon)$ of $\clW$
with respect to the maximum error probability is
\be
n^*_{\rm max}(\mathcal{W},\epsilon) := \inf 
\left\{ n \geq 1 : \inf_{(C_n,d_n)} \max \Big\{ 
P^{(n)}_{e,\max}, \frac{1}{|C_n|} \Big\} \leq \epsilon \right\}.
\label{def:scmax}
\ee

The point of these definitions is simple: Knowing 
$n^*_{\rm avg}(\mathcal{W},\epsilon)$ or $n^*_{\max}(\mathcal{W},\epsilon)$,
or having good bounds on them, allows the practitioner 
to estimate how large the blocklength needs to be 
so that explicit performance guarantees can be provided 
for both the rate and the error probability.

\subsection{Outline of main results}
\label{s:outline}

Section~\ref{sec:prelim} contains
the definitions, notation and terminology 
used throughout the paper, together with
some of the auxiliary results needed
in our analysis. Our first result is 
Theorem~\ref{thm:lecam}, where we adopt 
the same hypothesis testing setting as in the
Polyanskiy--Poor--Verd\'{u} meta-converse~\cite{PPV:10}
and prove a new one-shot channel-coding converse
which can be viewed
as a channel-coding version of the lower
bound in Le Cam's hypothesis testing lemma~\cite{lecam:86}.

In Section~\ref{sec:scmaxavg} we establish a simple
general relationship between $n^*_{\rm avg}(\clW,\epsilon)$ 
and $n^*_{\max}(\clW,\epsilon)$ for any channel~$\clW$:
Proposition~\ref{prop:scmaxavg}
states that we always have
$$n^*_{\rm avg}(\mathcal{W},\epsilon)
\leq n^*_{\rm max}(\mathcal{W},\epsilon) 
\leq n^*_{\rm avg}\Big(\mathcal{W},\frac{\epsilon}{2}\Big).$$

Our first main results are derived in Section~\ref{sec:scDMC}.
For any DMC $\clW$, Theorem~\ref{scDMC} states that,
$$n^*_{\rm avg}(\mathcal{W},\epsilon) \asymp 
\frac{\log( 1/\epsilon)}{C_{1/2}(\clW)},$$
where $C_{1/2}(\clW)$ is the channel's Augustin--Csisz\'{a}r
capacity of order $1/2$;
see Section~\ref{sec:prelim} for more details.
[As usual, 
for two nonnegative expressions $f$ and $g$ we write 
$f \asymp g$ to signify that there are absolute positive 
constants $c$ and $c'$ such that $cg \leq f \leq c' g$.]
In fact, Theorem~\ref{scDMC} establishes the following explicit
bounds,
\be
\frac{\log(1/\epsilon)- 2}{C_{1/2}(\clW)} 
\leq n^*_{\rm avg}(\mathcal{W},\epsilon) 
\leq \frac{2\log(1/\epsilon)}{C_{1/2}(\clW)} + 1,
\label{eq:DMCintro}
\ee
valid {\em uniformly} for all $\epsilon\in(0,1)$ and
all DMCs $\clW$.
Combined with Proposition~\ref{prop:scmaxavg}
these immediately yield
corresponding
bounds for $n^*_{\rm max}(\clW,\epsilon)$.

Qualitatively, Theorem~\ref{scDMC} shows that
the sample complexity scales like the logarithm of the
reciprocal of the error probability, and that the key 
property of the channel $\clW$ that determines
the sample complexity is not the usual capacity, but
its Augustin--Csisz\'{a}r capacity 
$C_{1/2}(\clW)$. 

The main tool in the proof of the upper bound in~\eqref{eq:DMCintro}
is the classical random coding bound.
For the lower bound,
we first relate the sample complexity
$n^*_{\rm avg}(\mathcal{W},\epsilon)$ to the more tractable
quantity
\be
N^*_{\rm avg}(\mathcal{W},\epsilon) 
:= \inf \left\{ n \geq 1 : \inf_{(C_n,d_n)}
\Big[
P^{(n)}_{e, {\rm avg}} + \frac{1}{|C_n|}\Big] \leq \epsilon \right\},
\label{eq:Nstar}
\ee
and then use
the Le Cam-style lower bound 
of Theorem~\ref{thm:lecam} to bound $N^*_{\rm avg}(\clW,\epsilon)$.

In Section~\ref{s:nonasympvsgaus}
we compare the sample complexity bounds 
of Theorem~\ref{scDMC}
with the corresponding estimates obtained
via the celebrated Gaussian-approximation 
approach to channel coding nonasymptotics.

Then in Section~\ref{s:actcom} we address
an important aspect of the sample complexity
framework. The optimal blocklength
$n^*_{\rm avg}(\clW,\epsilon)$ guarantees
the existence of codes that achieve
an error probability no greater than~$\epsilon$, 
but the rate of these codes
is only implicitly described: It
follows from the definitions that this
{\em actual} rate
is essentially given by:
$$
R_{\rm act}(\epsilon)
=\frac{\log(1/\epsilon)}{n^*_{\rm avg}(\clW,\epsilon)}.
$$
In Theorem~\ref{thm:limitingactualrate} we show that, 
under appropriate
conditions, for small $\epsilon$ 
the actual rate
$R_{\rm act}(\epsilon)$ is approximately
equal to the 
{\em Chernoff rate} $R_*=R_*(\clW)$ of the channel $\clW$,
defined as the unique fixed point of the reliability
function, $R_*=E^*(R_*)$. 

The sample complexity formulation 
of the channel coding problem 
via the sample complexities
$n^*_{\rm avg}(\clW,\epsilon)$ and
$n^*_{\rm max}(\clW,\epsilon)$,
therefore, enforces a ``hard'' restriction:
The error probability is no greater than $\epsilon$,
but there is no freedom in specifying a target
transmission rate -- beyond the guarantee that it is
given by $R_{\rm act}(\epsilon)$.
In Section~\ref{s:scassym}
we describe two different
extensions that allow for more general target
rates, and present preliminary results in that
direction.

Finally, in Section~\ref{sec:scgauss}, we establish
analogous sample complexity bounds for the
additive white Gaussian noise channel (AWGN) with
a quadratic power constraint. An important difference
from the case of DMCs is that here we need to establish
strong, nonasymptotic bounds on the error probability,
with an explicit, computable prefactor. 
For example, in Theorem~\ref{thm:probAWGN2}
we show that the best achievable error probability
$\epsilon^*(n,R,\sigma^2,P)$ at blocklength $n$ and
rate $R$ (in nats) for the AWGN channel with noise power $\sigma^2$
and input power constraint $P$ satisfies,
$$\epsilon^*(n,R,\sigma^2,P)\leq \Big[1+2\sqrt{
1+16(1+R)e^{2R}}\Big]e^{-nE_r(R,\gamma)},$$
where $\gamma=P/\sigma^2$ is the 
signal-to-noise ratio (SNR) and $E_r(R,\gamma)$
is the classical random coding exponent.

\section{Preliminaries for DMCs}
\label{sec:prelim}

As described earlier, a DMC $\clW=\{W^n\}$ from $A$ to $B$
is a sequence
of channels such that,
for all $n\geq 1$ and every pair of input and output sequences
$x^n=(x_1,\ldots,x_n)\in A^n$ and
$y^n=(y_1,\ldots,y_n)\in B^n$,
$$W^n(y^n|x^n)=\prod_{i=1}^n W(y_i|x_i),$$
where $W$ is a given collection of conditional
probability mass functions (p.m.f.s) $W(y|x)$,
$x\in A$, $y\in B$. Throughout, we assume 
that the input and output alphabets of a DMC
are finite sets. 

\medskip

\noindent
{\bf N.B.} 
Throughout Sections~\ref{sec:prelim}--\ref{sec:scDMC},
$\log=\log_2$ denotes the logarithm to base~2,
and all familiar information-theoretic functionals are
expressed in bits. In Section~\ref{sec:scgauss} on the AWGN 
channel, we use the natural logarithm $\log_e$,
and all rates and information-theoretic functionals
are expressed in nats.

\subsection{Error exponents}
\label{s:errorexp}

We recall the classical 
random coding upper bound for the error
probability of a DMC~\cite[Theorem~5.6.2]{gallager:book}.
First, 
let $\clW=\{W^n\}$ be a DMC from $A$ to $B$, 
and
recall the definition of the
best achievable average error probability
$\epsilon^*(n,R)$ in~\eqref{eq:epsilonstar}.
For any input p.m.f.\ $P$ on $A$,
Gallager's $E_0$ function~\cite{gallager:book} is,
$$E_0(\rho,P) := - \log \left( \sum_{y \in B} \Big( 
\sum_{x \in A} P(x) W(y|x)^{\frac{1}{1+\rho}} 
\Big)^{1+\rho} \right)\quad
\rho\geq 0.$$

\begin{proposition}[Random coding upper bound]
\label{prop:rcu}
Let $\clW$ be a DMC from $A$ to $B$.
For any blocklength $n\geq 1$,
any rate $0<R<\log|A|$, and any input p.m.f.\ $P$, 
we have
$$\epsilon^*(n,R)\leq 2^{-n \left( E_0(\rho,P) - \rho R \right)},
\qquad\mbox{for any}\;\rho\in[0,1].$$
\end{proposition}

Taking $\rho=1$ and optimizing over the input distribution,
we obtain the {\em cutoff rate} of the 
channel $\clW$~\cite{wozencraft:66}:
$$R_0(\clW):=\sup_{P} E_0(1,P).$$

\begin{corollary}[Cutoff rate]
Under the assumptions of Proposition~\ref{prop:rcu},
we always have:
$$\epsilon^*(n,R)\leq 2^{-n ( R_0(\clW) - R )}.$$
\end{corollary}

Since the best random coding exponent,
$$E_r(R):=\sup_P\sup_{\rho\in[0,1]}[E_0(\rho,P)-\rho R],$$
can be no better than the
reliability $E^*(R)$, the corollary
leads to the simple lower bound
$E^*(R)\geq R_0(\clW)-R$.
Due to this and numerous other useful properties,
the cutoff rate has 
been variously proposed as a convenient
benchmark for reliable 
communication~\cite{wozencraft:66,narayan:94,massey:74,blahut:10}.
For example, $R_0(\clW)$ plays a key role in the evaluation
of the maximum rate at which the average 
computational complexity of sequential decoding is bounded~\cite{arikan:88}, 
and in the characterization of achievable rates 
for convolutional codes~\cite{viterbi:67}.

\subsection{R\'{e}nyi divergence and capacities}
\label{s:Dalpha}

For any two p.m.f.s $P,Q$ on the
same discrete alphabet $A$, the {\em R\'{e}nyi divergence of order 
$\alpha \in (0,1)$} between $P$ and $Q$ 
is defined as~\cite{vanerven:14},
$$D_\alpha(P\|Q) := \frac{1}{\alpha-1}\log\left(\sum_{x \in A}P(x)^\alpha Q(x)^{1-\alpha} \right).$$
A pair of well-known inequalities~\cite{lecam:86,viaud:26ITT} 
relate $D_{1/2}(P\|Q)$ to the total variation distance
$\|\cdot\|_{\rm TV}$.

\begin{proposition}[R\'{e}nyi divergence and total variation]
\label{prop:DTV}
For any pair of p.m.f.s $P,Q$ on the same alphabet:
$$2^{-D_{1/2}(P\|Q)-1} \leq 1 - \|P-Q\|_{\rm TV} 
\leq 2^{-\frac{1}{2}D_{1/2}(P\|Q)}.$$
\end{proposition}

Recall that Shannon's classical formula
for a DMC $\clW=\{W^n\}$ from $A$ to $B$ states 
that its capacity $C(\clW)$ is given by
\begin{align}
C(\clW)
&=
	\sup_{X\sim P} I(X;Y)
	\label{eq:shannon1}\\
&=
	\sup_{P} \inf_Q D(PW\|P\times Q)
	\label{eq:shannon2}\\
&=
	\sup_{P} \inf_Q \BBE \big[D\big(W(\cdot|X)\|Q\big)\big].
	\label{eq:shannon3}
\end{align}
In~\eqref{eq:shannon1} the supremum is over all input p.m.f.s $P$ on $A$,
$X$ has marginal distribution $P$, the conditional distribution
of $Y$ given $\{X=x\}$ is $W(\cdot|x)$, and $I(X;Y)$ denotes the mutual
information. In~\eqref{eq:shannon2} and~\eqref{eq:shannon3}
the suprema are over all
input p.m.f.s $P$ on $A$, the infima are over all 
output p.m.f.s $Q$ on $B$,
$D(\cdot\|\cdot)$ denotes the relative entropy, and $PW$ denotes the joint
distribution of $X,Y$ as in~\eqref{eq:shannon1}.

In the same notation, 
replacing the relative entropy in~\eqref{eq:shannon2} with the 
R\'{e}nyi divergence, leads to the definition
of the {\em Sibson mutual information of order $\alpha$},
$$I^{\rm S}_{\alpha}(X;Y) := \inf_Q D_{\alpha} (PW\|P \times Q),$$
and the corresponding {\em Sibson capacity
of order $\alpha$}~\cite{sibson:69} 
$$C^{\rm S}_{\alpha}(\clW) := \sup_{X\sim P} I^{\rm S}_{\alpha}(X;Y),$$
for $\alpha\in(0,1)$.
Similarly, 
replacing the relative entropy in~\eqref{eq:shannon3} with the 
R\'{e}nyi divergence, leads to the definition
of the {\em Augustin--Csisz\'{a}r mutual information 
of order $\alpha$},
$$I^{\rm AC}_{\alpha}(X;Y) := \inf_Q \mathbb{E}\big[ D_{\alpha} \big(W(\cdot |X)
 \| Q \big) \big],$$
and the corresponding {\em Augustin--Csisz\'{a}r capacity
of order $\alpha$}~\cite{augustin:78},
$$C^{\rm AC}_{\alpha}(\clW) := \sup_{X\sim P} I^{\rm AC}_{\alpha}(X;Y),$$
for $\alpha\in(0,1)$.

The following relationships between
$C_\alpha^{\rm S}(\clW)$, $C_\alpha^{\rm AC}(\clW)$ and the 
cutoff rate $R_0(\clW)$ are established in~\cite{verdu:21}
and the references therein.

\begin{proposition}[Capacities and cutoff rate]
\label{prop:halfcap}
Let $X,Y$ be discrete random variables such that $X\sim P$ on $A$ 
and $Y$ given $\{X=x\}$ is distributed according to $W(\cdot|x)$ on $B$,
where $\clW=\{W^n\}$ is an arbitrary DMC from $A$ to $B$. Then:
\begin{itemize}
\item[$(i)$]
For any input distribution $P$ and any $\alpha\in(0,1)$, we have
$I^{\rm S}_{\alpha}(X;Y) \leq I^{\rm AC}_{\alpha}(X;Y).$
\item[$(ii)$]
For all $\alpha\in(0,1)$, we have
$C^{\rm S}_{\alpha}(\clW) = C^{\rm AC}_{\alpha}(\clW).$
\item[$(iii)$]
For all $\rho > 0$, we have
$$\rho  C^{\rm S}_{\frac{1}{1+\rho}}(\clW) 
= \rho C^{\rm AC}_{\frac{1}{1+\rho}}(\clW) = 
\sup_P E_0(\rho,P).$$
In particular, for $\rho = 1$,
\be
C^{\rm S}_{1/2}(\clW) = C^{\rm AC}_{1/2}(\clW) = R_0(\clW) = \sup_P
\left[
 - \log \left( \sum_{y \in B} \Big( \sum_{x \in A} 
P(x) \sqrt{W(y|x)} \Big)^2 \right)
\right].
\label{eq:identity}
\ee
\end{itemize}
\end{proposition}

In view of~\eqref{eq:identity}, we henceforth
write $C_{1/2}(\clW)$ for the common value of 
$C^{\rm S}_{1/2}(\clW)$, $C^{\rm AC}_{1/2}(\clW)$ and $R_0(\clW)$
for a given DMC $\clW$.
Similarly, we 
write $C_\alpha(\clW)$ for the common
value of $C_{\alpha}^{\rm AC}(\clW)=C_\alpha^{\rm S}(\clW)$
for all $\alpha\in(0,1)$.

\section{Properties of general channels}
\label{sec:scmaxavg}

Recall 
the definitions of the sample complexities
$n^*_{\rm avg}(\mathcal{W},\epsilon)$
and $n^*_{\rm max}(\mathcal{W},\epsilon)$
for a given channel $\clW=\{(W_{Y^n|X^n})\}$
from~\eqref{def:scavg} and~\eqref{def:scmax}.
Throughout, we assume that the input and output
alphabets $A,B$ of a general channel $\clW$
are Polish spaces equipped with their Borel
$\sigma$-algebras $\clA$ and $\clB$, respectively.
In this section we establish a general relationship 
between
$n^*_{\rm avg}(\mathcal{W},\epsilon)$
and $n^*_{\rm max}(\mathcal{W},\epsilon)$,
for arbitrary channels $\clW$, and we also 
derive a new general channel coding 
converse bound in terms of the
risk function in the definition of 
$n^*_{\rm avg}(\mathcal{W},\epsilon)$.

First, we observe that
$n^*_{\rm avg}(\mathcal{W},\epsilon)$
and $n^*_{\rm max}(\mathcal{W},\epsilon)$
are always tightly coupled.

\begin{proposition}[Maximal versus average sample complexity]
\label{prop:scmaxavg}
For an arbitrary channel $\mathcal{W}=\{(W_{Y^n|X^n})\}$ between
two general alphabets $A$ and $B$, and 
for any $\epsilon>0$,
$$n^*_{\rm avg}(\mathcal{W},\epsilon) 
\leq n^*_{\rm max}(\mathcal{W},\epsilon) 
\leq n^*_{\rm avg}\Big(\mathcal{W},\frac{\epsilon}{2}\Big),$$
with the obvious understanding that 
$n^*_{\rm max}(\mathcal{W},\epsilon)=
n^*_{\rm avg}(\mathcal{W},\epsilon)=1$
for all $\epsilon\geq 1$.
\end{proposition}

\noindent
{\sc Proof.}
The first inequality
is immediate from 
the obvious fact that 
$P^{(n)}_{e,{\rm avg}} \leq P^{(n)}_{e,\max}$
for all codes.
For the second inequality, it suffices
to show that
\be
\inf_{(C_n,d_n)} \max \Big\{ P^{(n)}_{e,\max}, \frac{1}{|C_n|} \Big\} 
\leq 2 \inf_{(C_n,d_n)} \max \Big\{P^{(n)}_{e,{\rm avg}}, \frac{1}{|C_n|} 
\Big\},
\label{eq:targetM}
\ee
which will follow from a standard argument.

Fix a blocklength $n \geq 1$ and a 
code $(C_n,d_n)$ with codebook $C_n=\{x^n(1),\ldots,x^n(M)\}$ of 
size $M=|C_n|$. 
We index the elements of 
$C_n$ in order of increasing error probability,
$$P^{(n)}_{e}(x^n(1)) \leq \cdots \leq P^{(n)}_{e}(x^n(M)),$$
and define a new codebook $C_n'=\{x^n(j):1\leq j\leq\lceil M/2\rceil\}$,
with associated decoder $d_n'$ which operates by applying $d_n$, 
keeping its decision if it belongs to $\{1,\ldots,\lceil M/2\rceil\}$, 
and otherwise returns~1.
Then $P_{e,\max}^{(n)}(C_n')\leq 2 P^{(n)}_{e,{\rm avg}}(C_n)$
while
$$\frac{1}{|C_n'|}=\frac{1}{\lceil \frac{M}{2} 
\rceil} \leq \frac{2}{M} 
=\frac{2}{|C_n|}.$$
Thus,
$$\max \Big\{ P^{(n)}_{e,\max}( C'_n ), \frac{1}{|C'_n|} \Big\} 
\leq 2 \max \Big\{ P^{(n)}_{e, {\rm avg}} ( C_n ), \frac{1}{|C_n|} 
\Big\}.$$ 
Since this construction applies to every original code $(C_n,d_n)$, 
taking the infimum over all such codes 
yields~\eqref{eq:targetM}
as desired.
\qed

In view of Proposition~\ref{prop:scmaxavg}, from now on
we concentrate on $n_{\rm avg}^*(\clW,\epsilon)$.

Next we derive a general
converse bound for channel coding,
using 
the same hypothesis testing framework as 
Polyanskiy--Poor--Verd\'{u}'s 
meta-converse~\cite{PPV:10,vazquez:13}.
 
Consider a code $(C_n,d_n)$ 
of blocklength $n$,
with a codebook
$C_n=(x^n(1),\ldots,x^n(M))$ of size $M=|C_n|$,
for 
an arbitrary channel $W_{Y^n|X^n}$ between
two general input and output alphabets $A$ and $B$.
Write $P_{C_n}$ for the p.m.f.\ on $A^n$
obtained by drawing codewords $c\in A^n$ from $C_n$
uniformly at random. For an arbitrary output distribution
$Q_n$ on $B^n$, consider the binary hypothesis test,
$$\mathcal{H}_0 : P_{C_n} W_{Y^n|X^n}\quad\mbox{versus}\quad
\mathcal{H}_1 : P_{C_n} \times Q_n,$$
where $P_{C_n} W_{Y^n|X^n}$ denotes the joint 
distribution of $(X^n,Y^n)$ when
$X_n\sim P_{C_n}$ and 
$W_{Y^n|X^n}(\cdot|\cdot)$
denotes the regular conditional distribution of
$Y^n$ given $X^n$.
A (possibly randomized) test is a map
$T : C_n \times B^n \rightarrow [0,1]$ which, 
for $(x^n,y^n) \in C_n \times B^n$, accepts $\mathcal{H}_1$ 
with probability $T(x^n,y^n)$, and accepts $\mathcal{H}_0$ otherwise. 
The type I and type II errors of $T$ are given by
\be
\mathbb{E}_{\clH_0}[T(X^n,Y^n)]\quad\mbox{and}
\quad \mathbb{E}_{\clH_1}[1-T(X^n,Y^n)],
\label{eq:test}
\ee
respectively. 
The optimal test is the Neyman--Pearson test,
which, given a constraint $\beta>0$ on the type II 
error probability, achieves the optimal
type I error probability, commonly denoted by
$\alpha_\beta (P_{C_n} W_{Y^n|X^n}, P_{C_n} \times Q_n)$.

The decoder $d_n$ can be used to define a test $T_{d_n}$ 
as follows. If $d_n$ is deterministic, let
$T_{d_n}(x^n(j),y^n)=\IND\{d_n(y^n)\neq j\}$, for
$x^n(j)\in C_n, y^n\in B^n$. More generally, 
if $d_n$ is possibly randomized, let
\be
T_{d_n}(x^n(j),y^n)=\BBP(d_n(Y^n)\neq j|X^n=x^n(j),Y^n=y^n),
\label{eq:randomized}
\ee
where the probability is only over the decoder's randomness.
It is easy to see that, then,
the error probabilities of $T_{d_n}$ satisfy,
\be
\mathbb{E}_{\clH_0}[T_{d_n}(X^n,Y^n)]=P^{(n)}_{e,{\rm avg}}\quad\mbox{and}
\quad \mathbb{E}_{\clH_1}[1-T_{d_n}(X^n,Y^n)]=\frac{1}{|C_n|}.
\label{eq:probs}
\ee
The meta-converse~\cite[Theorems~26-27]{PPV:10}
is the observation that for this test we must necessarily 
have
$$P^{(n)}_{e,{\rm avg}} \geq 
\alpha_{\frac{1}{|C_n|}} \big( 
P_{C_n} W_{Y^n|X^n} , P_{C_n} \times Q_n \big).$$

We use this setting to prove a different converse,
in the spirit of Le Cam's lemma~\cite{lecam:86},
which is better suited for our purposes.

\begin{theorem}[Le Cam's lower bound for channel coding]
\label{thm:lecam}
Suppose $(C_n,d_n)$ is an arbitrary code of blocklength $n\geq 1$
for an arbitrary channel $W_{Y^n|X^n}$ from $A$ to $B$. 
Then,
\begin{align}
P^{(n)}_{e,{\rm avg}} + \frac{1}{|C_n|} 
&\geq 
	\sup_{Q_n} \left[ 1 - \left\| P_{C_n} W_{Y^n|X^n} 
	- P_{C_n} \times Q_n \right\|_{\rm TV} \right] 
	\label{eq:LC1}\\
&= 
	\sup_{Q_n} \left[ 1 - \frac{1}{|C_n|} 
	\sum_{x^n\in C_n}
	\left\| W_{Y^n|X^n}(\cdot|x^n) - Q_n \right\|_{\rm TV}
	\right],
	\label{eq:LC2}
\end{align}
where the supremum is over all output distributions $Q_n$ on $B^n$.
\end{theorem}

\noindent
{\sc Proof.}
For any distribution $Q_n$ on $B^n$, consider the binary 
hypothesis test~\eqref{eq:test} and define the test
$T_{d_n}$ as in~\eqref{eq:randomized}. Using~\eqref{eq:probs}
we have
\begin{align*}
P^{(n)}_{e, {\rm avg}} + \frac{1}{|C_n|} 
&= 
	\mathbb{E}_{\clH_0}[T_{d_n}(X^n,Y^n)] 
	+ \mathbb{E}_{\clH_1}[1-T_{d_n}(X^n,Y^n)]\\
&\geq 
	\inf_{{\rm all\; tests}\;T}
	\Big[
	\mathbb{E}_{\clH_0} [T(X^n,Y^n)] + \mathbb{E}_{\clH_1}[1-T(X^n,Y^n)]
	\Big],
\end{align*}
and applying Le Cam's hypothesis testing lemma~\cite{lecam:86}, 
$$
P^{(n)}_{e, {\rm avg}} + \frac{1}{|C_n|} \geq
1 - \left\| P_{C_n} W_{Y^n|X^n} - P_{C_n} \times Q_n \right\|_{\textup{TV}}.
$$
Since the left-hand side is independent of $Q_n$, 
the bound~\eqref{eq:LC1} follows by taking the supremum of both
sides over all $Q_n$.
The expression~\eqref{eq:LC2} is an immediate consequence
of the definitions, since 
$$\left\| P_{C_n} W_{Y^n|X^n} - P_{C_n} \times Q_n \right\|_{\textup{TV}}
=
	\frac{1}{|C_n|}\sum_{x^n\in C_n}
	\left\| W_{Y^n|X^n}(\cdot|x^n) - Q_n \right\|_{\rm TV},$$
for every $Q_n$.
\qed

\section{Sample complexity of DMCs}
\label{sec:scDMC}

\subsection{Nonasymptotic bounds}
\label{s:DMCbounds}

Our first main result gives explicit upper and lower bounds
on the channel coding sample complexities
$n^*_{\rm avg}(\mathcal{W},\epsilon)$
and
$n^*_{\rm max}(\mathcal{W},\epsilon)$ 
defined in~\eqref{def:scavg} and~\eqref{def:scmax}
for any DMC $\clW$.

\begin{theorem}[Sample complexity of DMCs]
\label{scDMC}
For any discrete memoryless channel $\mathcal{W}$ with 
Augustin--Csisz\'{a}r capacity $C_{1/2}(\clW) > 0$, the maximum
and average channel coding sample complexities of $\mathcal{W}$ satisfy, 
for all $\epsilon \in (0,1)$,
\begin{align}
\frac{\log(1/\epsilon) - 2}{C_{1/2}(\clW)} 
&\;\leq\; n^*_{\rm avg}(\mathcal{W},\epsilon) 
\;\leq\; \frac{2\log(1/\epsilon)}{C_{1/2}(\clW)} + 1,
\label{dmc:1}\\
\frac{\log(1/\epsilon) - 2}{C_{1/2}(\clW)} 
&\;\leq\; n^*_{\rm max}(\mathcal{W},\epsilon) 
\;\leq\; \frac{2\big(\log(1/\epsilon) + 1\big)}{C_{1/2}(\clW)} + 1.
\label{dmc:2}
\end{align}
In particular, for $0 < \epsilon < \min \big\{ \frac{1}{8}, 
\frac{1}{\min \{ |A|,|B| \}} \big\}$, we have,
\begin{align}
\frac{\log(1/\epsilon)}{3C_{1/2}(\clW)} 
&\;\leq \;
n^*_{\rm avg}(\mathcal{W},\epsilon) 
\;\leq\;
\frac{3\log(1/\epsilon)}{C_{1/2}(\clW)},
\label{dmc:3}\\
\frac{\log(1/\epsilon)}{3C_{1/2}(\clW)} 
&\;\leq \;
n^*_{\rm max}(\mathcal{W},\epsilon) 
\leq \frac{5\log(1/\epsilon)}{C_{1/2}(\clW)}.
\label{dmc:4}
\end{align}
\end{theorem}

\noindent
{\sc Proof.}
The second pair of bounds in~(\ref{dmc:3}) and~(\ref{dmc:4}) 
follow from~(\ref{dmc:1}) and~(\ref{dmc:2}) by direct calculation,
using the elementary fact that $C_{1/2}(\clW) \leq \min \{
\log|A|,\log|B| \}$. Also,~(\ref{dmc:2}) follows directly 
from~(\ref{dmc:1}) combined with Proposition~\ref{prop:scmaxavg}. 
So it remains to prove~(\ref{dmc:1}).

Let
$\epsilon(n)$ denote the risk function
in the definition of $n^*_{\rm avg}(\clW,\epsilon)$,
$$\epsilon(n) := \inf_{(C_n,d_n)}
\max \left\{ P^{(n)}_{e,{\rm avg}} , \frac{1}{|C_n|} \right\}.$$
Given $\epsilon \in (0,1)$,
using the random coding upper bound in
Proposition~\ref{prop:rcu} we have,
\begin{align}
\epsilon(n) 
&= 
	\inf_{0 < R \leq \log|A|}\;
	\inf_{(C_n,d_n)\,:\, |C_n| = \lceil 2^{nR} \rceil}
	\max \left\{ P^{(n)}_{e,{\rm avg}} , \frac{1}{\lceil 2^{nR}\rceil}
	\right\} 
	\label{dmcub:1} \\
&\leq 
	\inf_{0 < R \leq \log|A|}
	\max \left\{ 2^{-n ( C_{1/2}(\clW) - R )}, 
	2^{-nR} \right\} 
	\nonumber\\
&\leq 
	2^{-n C_{1/2}(\clW)/2}.
	\nonumber
\end{align}
The upper bound in~\eqref{dmc:1}
follows by the definition of
$n^*_{\rm avg}(\mathcal{W},\epsilon)$.

For the lower bound,
let $\epsilon'(n)$ denote the risk function
corresponding to
$N^*_{\rm avg}(\mathcal{W},\epsilon)$ 
defined in~\eqref{eq:Nstar}, i.e.,
$$
\epsilon'(n)
=\inf_{(C_n,d_n)}\Big[
P^{(n)}_{e,{\rm avg}}
+ \frac{1}{|C_n|}
\Big].$$
By the definitions we have that
$\epsilon'(n) \leq 2 \epsilon(n)$,
and hence, for all $\epsilon$,
\be
n^*_{\rm avg}(\clW,\epsilon)
\geq
N^*_{\rm avg}(\clW,2\epsilon).
\label{eq:nstarN}
\ee

Using the Le Cam lower bound 
	in Theorem~\ref{thm:lecam}
	followed by the R\'{e}nyi divergence
	inequality in Proposition~\ref{prop:DTV},
	for any code $(C_n,d_n)$,
	we have,
\begin{align*}
P^{(n)}_{e,{\rm avg}}
+ \frac{1}{|C_n|}
	&\geq 
		\sup_{Q_n} \frac{1}{|C_n|} 
		\sum_{x^n\in C_n}\Big[ 1 - \| W^n(\cdot|x^n)-Q_n
		\|_{\rm TV}\Big]
		\\
	&\geq
		\sup_{Q_n} \frac{1}{|C_n|} 
		\sum_{x^n\in C_n} \exp\big\{-D_{1/2}\big(W^n(\cdot|x^n) 
		\big\| Q_n\big)-1\big\},
\end{align*}
	where $\exp\{a\}$ denotes $2^a$ for any expression $a$,
	and the supremum in both cases
	is over
	all p.m.f.s $Q_n$ on $B^n$.
	Restricting to product p.m.f.s $Q_n$ 
	of the form $Q^n$, using the tensorization
	property of R\'{e}nyi divergence, and 
	then applying Jensen's inequality 
	with respect to the convex map $x\mapsto 2^{-x-1}$,
	$x\in\RL$, gives
\begin{align*}
P^{(n)}_{e,{\rm avg}}
+ \frac{1}{|C_n|}
	&\geq
		\sup_Q \frac{1}{|C_n|} 
		\sum_{x^n\in C_n}\exp\big\{
			-D_{1/2}\big(W^n(\cdot|x^n) 
			\big\| Q^n\big)-1
					\big\}\\
	&= 
		\sup_Q 
		\sum_{x^n\in C_n}
		\frac{1}{|C_n|} 
		\exp\Big\{- \sum_{i=1}^n 
		D_{1/2}\big(W(\cdot|x_i)\big\|Q\big)-1\Big\} 
		\\
	&\geq 
		\sup_Q \exp\Big\{-n
		\Big[ \frac{1}{n|C_n|} \sum_{x^n\in C_n} 
		\sum\limits_{i =1}^n D_{1/2}
		\big(W(\cdot|x_i)\big\|Q\big) 
		\Big]
		-1\Big\}.
\end{align*}
Let $\hat{P}$ denote the empirical distribution
on $A$ induced by uniformly choosing a codeword
$x^n\in C_n$:
$$\hat{P}(z)=\frac{1}{n|C_n|}\sum_{x^n\in C_n}
[\#\;\mbox{times $z$ appears in $x^n$}].$$
Writing 
$\hat{X}\sim \hat{P}$, the last bound above can be expressed as
$$
P^{(n)}_{e,{\rm avg}}
+ \frac{1}{|C_n|}
\geq
\sup_Q \exp\Big\{-n \mathbb{E} 
	\big[ D_{1/2}\big(W(\cdot|\hat{X})\big\|Q\big) \big] -1\Big\}.
$$
and taking the infimum over all codes $(C_n,d_n)$ yields,
$$
\epsilon'(n)
    \geq \exp\Big\{-n \sup_P\inf_Q \mathbb{E}
	\big( D_{1/2}(W(\cdot|X)\big\|Q\big) \big) -1\Big\} 
    = 2^{-n C_{1/2}(\clW) -1},
$$
where the inequality follows 
by replacing the
infimum over all codes by the infimum over 
the larger class of all
input distributions $P$ on $A$, 
and the equality follows from the definition
of the Augustin--Csisz\'{a}r capacity.
Therefore, 
$$N_{\rm avg}^*(\clW,\epsilon)\geq\frac{\log(1/\epsilon)-1}{C_{1/2}(\clW)},
$$
and the lower bound in~\eqref{dmc:1} follows by~\eqref{eq:nstarN},
completing the proof.
\qed

\noindent
{\bf Remarks.}
\begin{enumerate}
\item {\em Proof.}
In the achievability part of the proof,
although the random coding bound 
is a tool stronger than necessary 
for the purpose of establishing the 
upper bounds in~\eqref{dmc:1}--\eqref{dmc:4},
as we will see the resulting argument easily 
generalizes
to the asymmetric formulation of 
the problem in Theorem~\ref{thm:asymmetric}.
The proof of the converse applies 
to any (not necessarily discrete)
memoryless channel, including in the case of 
input power constraints; see,
e.g., the proof of Theorem~\ref{scGauss} 
in Section~\ref{s:awgnnstar}.

\item {\em Augustin--Csisz\'{a}r capacity.}
The results in Theorem~\ref{scDMC} can be naturally
viewed as achievability and converse channel-coding results. 
Interestingly, the key information-theoretic functional 
that determines optimal performance in this setting
is not the usual Shannon capacity of the channel $C(\clW)$,
but its Augustin--Csisz\'{a}r capacity $C_{1/2}(\clW)$ 
of order $1/2$.

\item {\em Sharper bounds.}
For many individual DMCs, in principle (as well as in practice)
it is possible to obtain even sharper numerical bounds
for $n^*_{\rm avg}(\clW,\epsilon)$
and $n^*_{\rm max}(\clW,\epsilon)$ than those
in the theorem, by numerically
evaluating any one
of the numerous one-shot achievability and converse
results known already; see, e.g.,~\cite{PPV:10,gao:26}.
But our point here is not to provide the sharpest
bounds possible. Rather, we wish to show that
the sample complexity can be {\em uniformly} bounded
over {\em all} $\epsilon\in(0,1)$ and {\em all} DMCs $\clW$,
through simple, informative, and easily accessible expressions.
The resulting bounds do not give the closest numerical
approximations for any one specific DMC, 
but they reveal the essential way in which sample 
complexity depends on the threshold $\epsilon$ 
(i.e., that it varies as $\log(1/\epsilon)$) 
and how it depends on the channel $\clW$,
which turns out to be only via the 
Augustin--Csisz\'{a}r capacity $C_{1/2}(\clW)$.

\item {\em The actual rate.}
In terms of channel coding,
the value of the fundamental limit $n^*_{\rm avg}(\clW,\epsilon)$
describes a blocklength
that guarantees the existence of codes with 
average error probability no greater than $\epsilon$ and with
rate essentially at least as large as
$$R_{\rm act}(\epsilon):=
\frac{\log(1/\epsilon)}{n^*_{\rm avg}(\clW,\epsilon)}.$$
This is hard to interpret at first glance,
but for small $\epsilon>0$ 
it is approximately
equal to what we call the {\em Chernoff rate} $R_*=R_*(\clW)$
of the channel $\clW$,
defined as the fixed point of the reliability function,
$R_*=E^*(R_*)$. An extensive discussion 
of $R_{\rm act}(\epsilon)$ and of $R_*$ is given
in Section~\ref{s:actcom}. 

\item {\em Different target rates.}
As can be seen from the definitions of the sample
complexity fundamental limits
$n^*_{\rm avg}(\clW,\epsilon)$ and
$n^*_{\rm max}(\clW,\epsilon)$,
the formulation of optimal channel coding
in terms of 
$n^*_{\rm avg}(\clW,\epsilon)$ and
$n^*_{\rm max}(\clW,\epsilon)$ implicitly
enforces an unnatural requirement: The
communication rate is essentially mandated
to be the same as the exponent of the
probability of error. This 
is both limiting and
artificial in terms of applications. 
In Section~\ref{s:scassym} we describe
how the sample complexity framework
and the bounds of Theorem~\ref{scDMC}
can be extended in various ways to incorporate
more general rate and error probability 
requirements.

\item {\em Zero rates.}
In the case when $C_{1/2}(\clW) = 0$, it is straightforward
to simply evaluate the sample complexity 
exactly from the definitions. Indeed, 
$C_{1/2}(\clW) = 0$ if and only if the channel 
input and output are independent,
in which case 
both 
the optimal average and optimal maximum error probabilities
equal $1 - 1/M$ 
for all codebooks of size~$M$.
Therefore,
assuming $|A|\geq 2$, we have
$n^*_{\rm avg}(\mathcal{W},\epsilon) = 
n^*_{\rm max}(\mathcal{W},\epsilon) = 1$ for $\epsilon\in[1/2,1]$,
and $n^*_{\rm avg}\left(\mathcal{W},\epsilon\right) = 
n^*_{\rm max}\left(\mathcal{W},\epsilon\right) = \infty$ 
for $0 < \epsilon < 1/2$. 
\end{enumerate}

\subsection{Sample complexity versus Gaussian approximation}
\label{s:nonasympvsgaus}

The most commonly used approach to obtaining useful
finite-blocklength information on optimal channel coding
is via normal approximation~\cite{strassen:64b,PPV:10},
leading to expansions like the one in~\eqref{eq:expansion1}
described in the Introduction. 
A key difference between the bounds 
in Theorem~\ref{scDMC} and these central limit theorem
(CLT)-based approximations
is that the sample complexity bounds are absolute and 
valid for any DMC $\clW$ and all $\epsilon\in(0,1)$,
whereas CLT-based expansions only offer approximations
without accuracy guarantees.
Also, Gaussian approximation is 
expected to be more accurate in the (CLT) regime 
corresponding to moderate values of 
the error probability $\epsilon$
and less accurate in the large deviations regime 
when $\epsilon$ is close to zero.
By contrast, 
the sample complexity bounds in~(\ref{dmc:1}) 
and~(\ref{dmc:2}) hold uniformly 
over all $\epsilon \in (0,1)$.

\begin{figure}[ht!]
\centering
\includegraphics[width=6.2in]{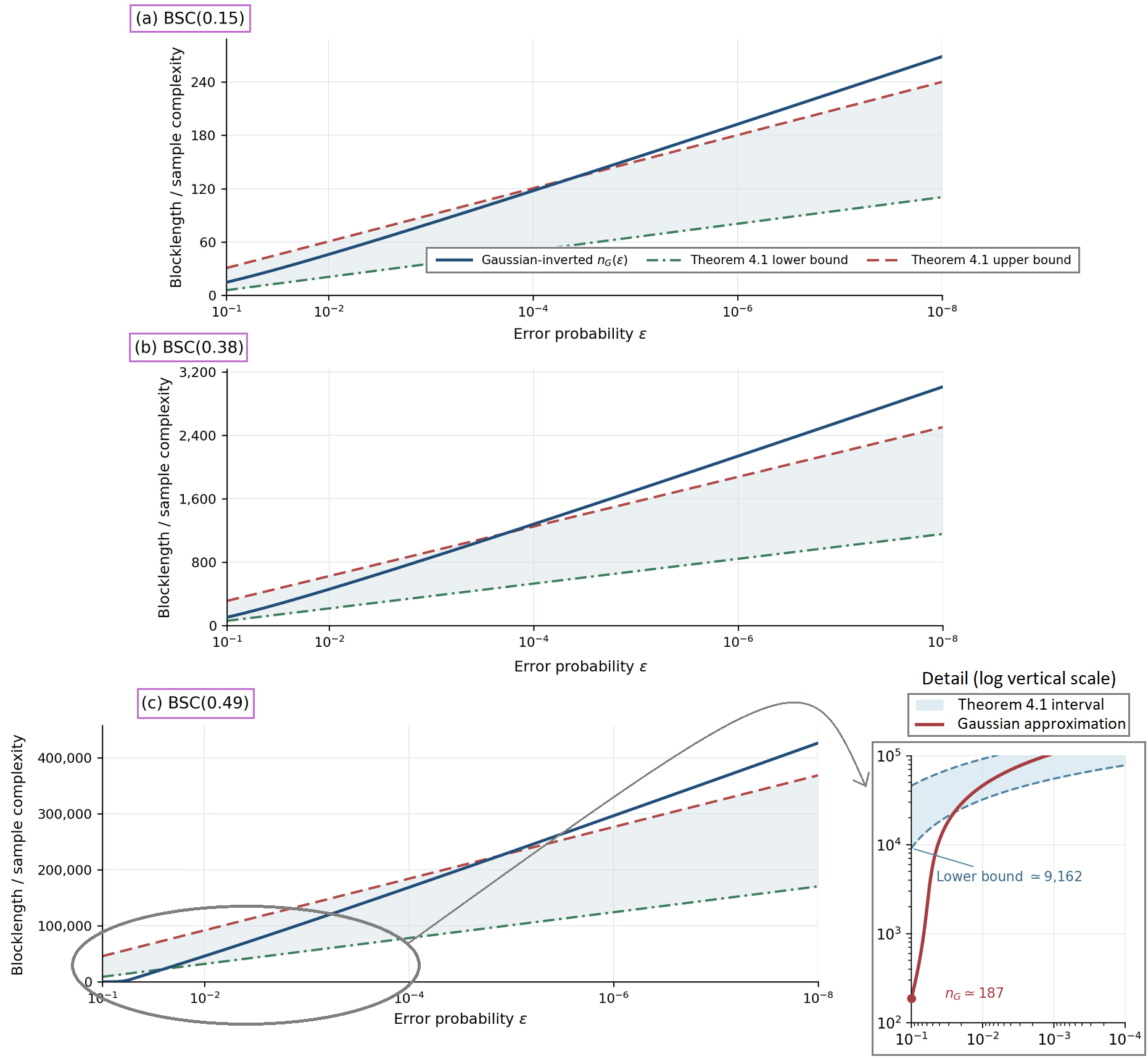}
\caption{Comparison of the Gaussian approximation 
$n_G(\epsilon)$ 
to $n_{\rm avg}^*(\mathcal{W},\epsilon)$
with the bounds in Theorem~\ref{scDMC},
in the case of the BSC with parameter:
$(a)$ $p=0.15$,
$(b)$ $p=0.38$, and 
$(c)$ $p=0.49$.}
\label{fig:example1}
\end{figure}

For an explicit numerical comparison, we evaluate 
these two approaches in the case of the binary
symmetric channel (BSC) with crossover probability $p$.
We plot the upper and lower bounds 
to $n^*(\mathcal{W},\epsilon)$ in (\ref{dmc:1}) as 
a function of $\epsilon$, as well as the Gaussian 
approximation $n_G(\epsilon)$ to 
$n^*(\mathcal{W},\epsilon)$ obtained as follows.
For $p\in(0,1/2)$, the optimal rate expansion in~\cite[Theorem~52]{PPV:10} 
leads to the approximation
\be
\log M^*(n,\epsilon) \approx \log M_G(n,\epsilon) 
:= n ( 1 - h(p) ) - \sqrt{np(1-p)} \log \Big(\frac{1-p}{p}\Big)
Q^{-1}(\epsilon) 
+ \frac{1}{2} \log n, 
\label{nonasympapprox:bsc}
\ee
where $h(p) = - p \log p - (1-p) \log \left( 1 - p \right)$ is the 
binary entropy function and $Q$ is the standard 
Gaussian tail function.
Inverting~(\ref{nonasympapprox:bsc}) 
by solving for $n$, we let $n_G(\epsilon)$ be the smallest $n$ 
such that $M_G(n,\epsilon)$ satisfies the sample-complexity 
codebook size constraint, i.e., 
$$\log M_G(n,\epsilon) \geq \log \frac{1}{\epsilon}.$$
The results, shown in Figure~\ref{fig:example1}, clearly indicate that
indeed the Gaussian approximation estimate $n_G(\epsilon)$ 
is significantly less accurate for smaller values of $\epsilon$.

\begin{figure}[ht!]
\centering
\includegraphics[width=3.9in]{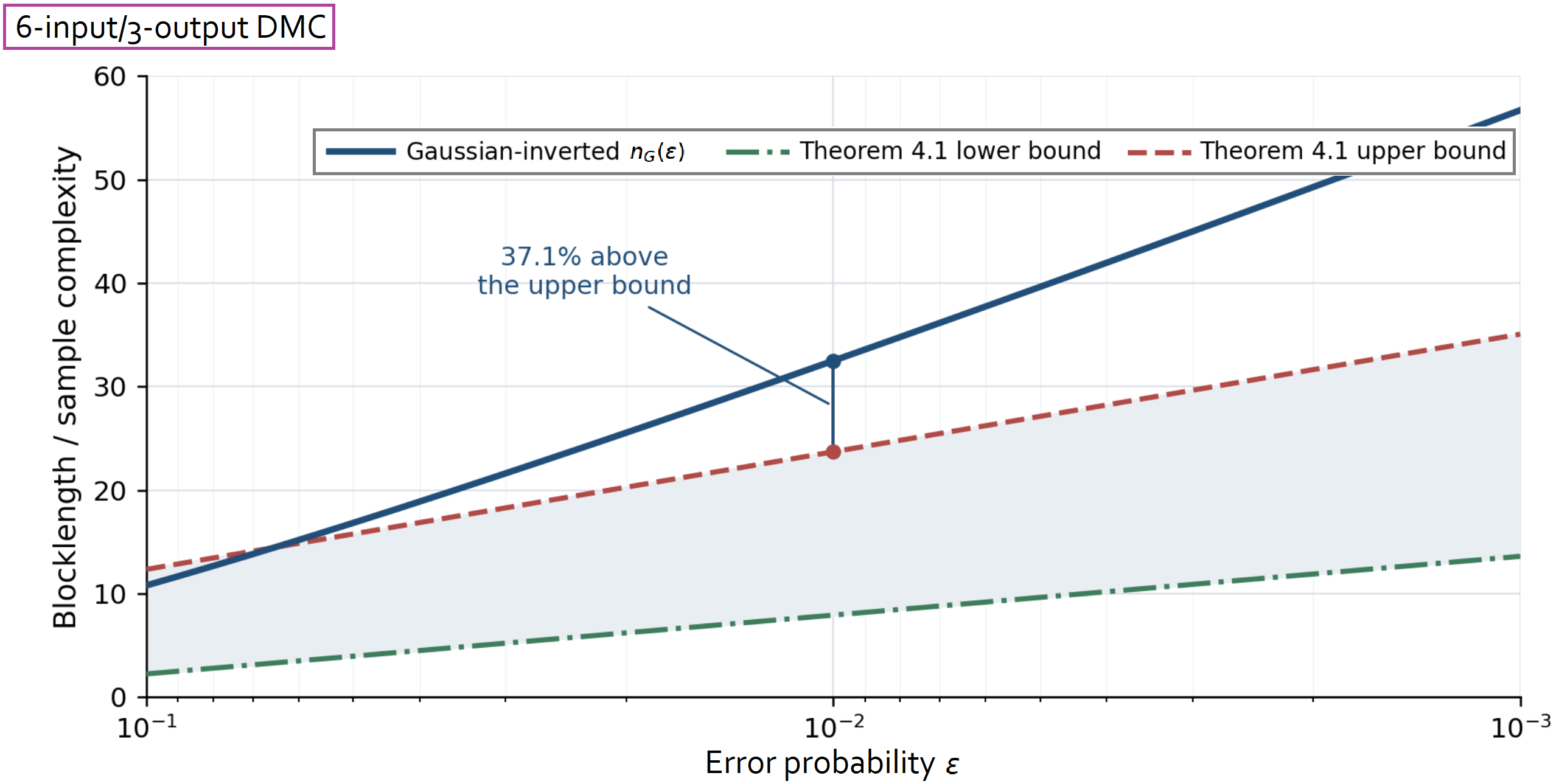}
\caption{Comparison of the Gaussian approximation 
$n_G(\epsilon)$ 
to $n_{\rm avg}^*(\mathcal{W},\epsilon)$
with the bounds in Theorem~\ref{scDMC},
in the case of the 6-input/3-output DMC
in~\eqref{eq:W}.}
\label{fig:example2}
\end{figure}

A simple example where the same phenomenon is even more pronounced
and at quite moderate values of $\epsilon$,
is for the 6-input/3-output DMC $\clW=\{W^n\}$ given by the transition 
matrix
\be
W=
\begin{pmatrix}
0&1/2&1/2\\
1/2&0&1/2\\
1/2&1/2&0\\
0.78&0.11&0.11\\
0.11&0.78&0.11\\
0.11&0.11&0.78
\end{pmatrix}.
\label{eq:W}
\ee
Once again, the corresponding CLT expansion for the optimal
rate in~\cite{polyanskiy:thesis,tomamichel:13}
leads to the approximation
\be
\log M^*(n,\epsilon) \approx \log M_G(n,\epsilon) 
:= nC(\clW) - \sqrt{nV(\clW)} 
Q^{-1}(\epsilon) 
+ \frac{1}{2} \log n, 
\label{eq:DMC6G}
\ee
where $C(\clW)\approx 0.605$ is the capacity
and $V(\clW)\approx 1.37$ is the dispersion.
Defining the Gaussian approximation $n_G(\epsilon)$ to
$n^*_{\rm avg}(\clW,\epsilon)$ as before but
using~\eqref{eq:DMC6G} in place of~\eqref{nonasympapprox:bsc},
the resulting values 
of $n_G(\epsilon)$ and of the bounds in
Theorem~\ref{scDMC}
are shown in Figure~\ref{fig:example2}.

\begin{figure}[ht!]
\centering
\includegraphics[width=4.3in]{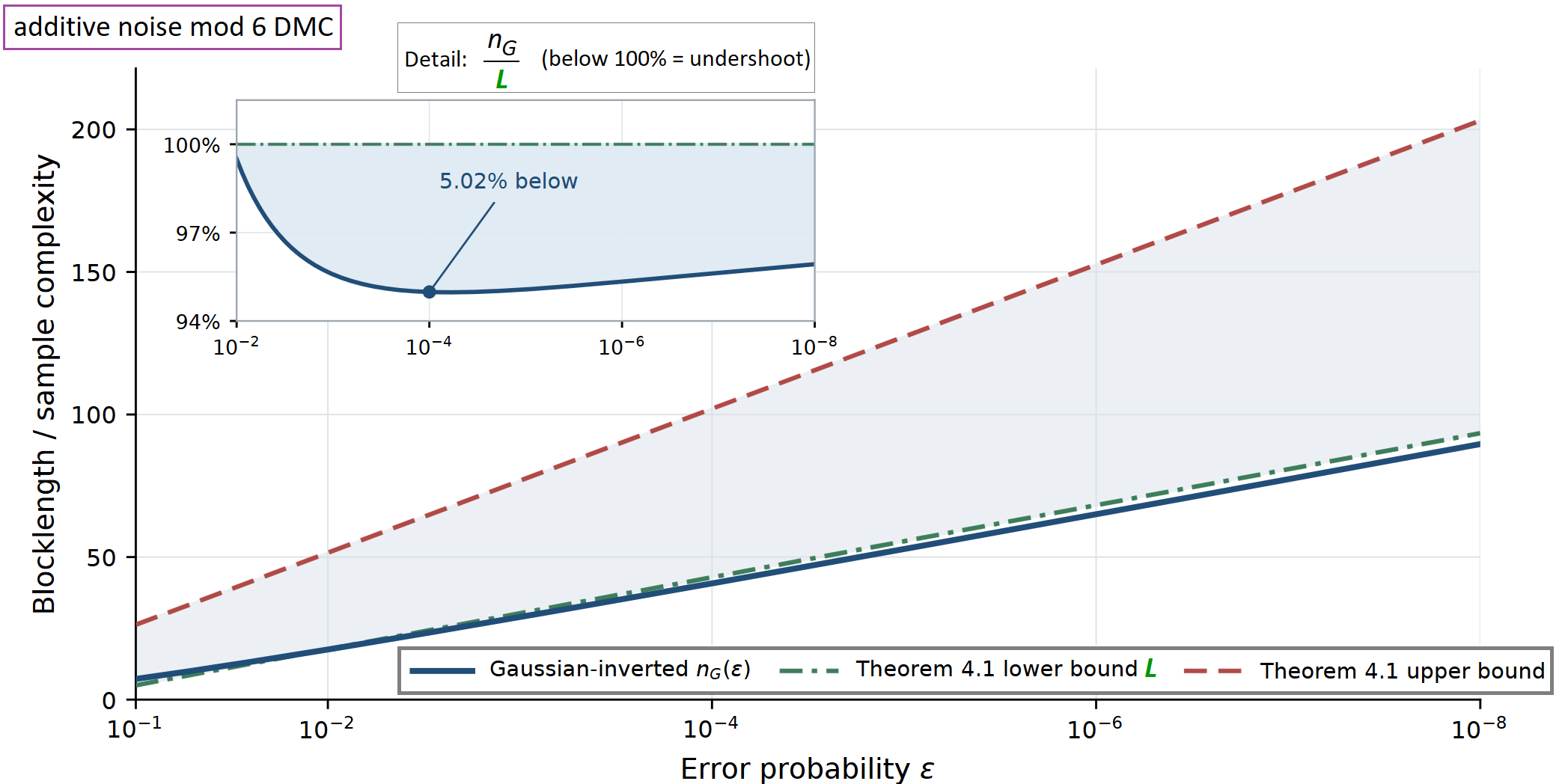}
\caption{Comparison of the Gaussian approximation 
$n_G(\epsilon)$ 
to $n_{\rm avg}^*(\mathcal{W},\epsilon)$
with the bounds in Theorem~\ref{scDMC},
in the case of the (mod 6)-additive noise
DMC with noise p.m.f.\ 
$P_Z=(0,0.201,0.199,1/5,1/5,1/5)$.}
\label{fig:example3}
\end{figure}

Considering the DMC defined by
$Y=X+Z\;({\rm mod}\; 6)$ on $A=B=\{0,1,2,3,4,5\}$,
where $Z$ is independent noise with p.m.f.\
$P_Z=(0,0.201,0.199,1/5,1/5,1/5)$,
shows that the Gaussian approximation
$n_G(\epsilon)$ defined via~\eqref{eq:DMC6G} 
as before~\cite{altug:14b}
may also 
substantially underestimate $n_{\rm avg}^*(\clW,\epsilon)$.
The results are shown in Figure~\ref{fig:example3}.

\subsection{The actual communication rate}
\label{s:actcom}

As noted in Remark~4 in Section~\ref{s:DMCbounds},
the value of the fundamental limit $n_{\rm avg}^*(\clW,\epsilon)$
describes a blocklength at which we can transmit information
through the channel $\clW$ from $A$ to $B$ 
with error probability no greater
than $\epsilon$, but the rate at which this can be done is
less explicit. In order to explore this further, 
we observe that, 
as in~\eqref{dmcub:1}
in the proof of Theorem~\ref{scDMC},
we can equivalently express $n^*_{\rm avg}(\clW,\epsilon)$ 
in terms
of the optimal error probability $\epsilon^*(n,R)$
defined in~\eqref{eq:epsilonstar}:
\be
n^*_{\rm avg}(\mathcal{W},\epsilon) 
= \inf \left\{ n \geq 1 : \inf_{0 < R \leq \log|A|} \max
\Big \{ \epsilon^*(n,R) , \frac{1}{\left\lceil 2^{nR} \right\rceil} 
\Big\} \leq \epsilon \right\}.
\label{scavgrate}
\ee
Since $\epsilon^*(n,R)$ is monotonic in $R$, the infimum 
in~\eqref{scavgrate} is essentially achieved 
(ignoring integer constraints) with equality.
Therefore, 
we define the \textit{actual communication rate} for a given $\epsilon > 0$ as:
$$R_{\text{act}}(\epsilon) :=  \frac{\log(1 / \epsilon)}{n^*_{\rm avg}
(\mathcal{W},\epsilon)}.$$
Let $R_*$ denote the unique fixed point of the
reliability function $E^*(R)$ of $\clW$ (see below).
Although $R_{\rm act}(\epsilon)$ is only implicitly defined 
via $n^*_{\rm avg}(\clW,\epsilon)$,
we will show that, under mild appropriate conditions,
$$R_{\rm act}(\epsilon)\approx R_*\quad\mbox{for}\;\epsilon>0\;\mbox{close
to}\;0.$$

Recall that the {\em reliability function} $E^*(R)$ of a channel $\clW$
is defined as the optimal error exponent,
$$E^*(R):=\limsup_{n\to\infty}-\frac{1}{n}\log\epsilon^*(n,R).$$
If $\clW$ is a DMC and its {\em zero-error capacity} $C_{\rm ZE}(\clW)$ 
is zero,
then $E^*(R)$ is finite,
nonincreasing, and continuous in $R\in(0,C(\clW))$, with
$$\lim_{R\downarrow 0} E^*(R)>0\quad\mbox{and}\quad 
\lim_{R\uparrow C} E^*(R)=0,$$
see, e.g.,~\cite{haroutunian:book}.
Therefore, 
whenever $C(\clW)>0$ and $C_{\rm ZE}(\clW)=0$, the reliability function
$E^*(R)$ has a unique fixed point $R_*=R_*(\clW)\in(0,C(\clW))$
such that $E^*(R_*)=R_*$. We call $R_*=R_*(\clW)$ 
{\em the Chernoff rate of $\clW$}.

Theorem~\ref{thm:limitingactualrate} is proved in Appendix~\ref{app:actual}.

\begin{theorem}[Limiting actual rate]
\label{thm:limitingactualrate}
Let $\clW$ be
any DMC with capacity $C(\clW)>0$ and
zero-error capacity $C_{\rm ZE}(\clW)=0$. 
\begin{itemize}
\item[$(a)$]
We  have,
\be
\limsup_{\epsilon \downarrow 0} 
R_{\rm act}(\epsilon) = R_*,
\label{eq:parta}
\ee
where $R_*=R_*(\clW)$ is the {\em Chernoff rate} of $\clW$,
defined as the unique fixed point of the reliability
function $E^*(R)$ of $\clW$. 
\item[$(b)$] 
If, moreover, the reliability function $E^*(R_*)$
exists as a limit at $R_*$, i.e.,
$$
E^*(R_*)= \lim_{n\to\infty}-\frac{1}{n}\log \epsilon^*(n,R_*),$$
then,
$$\lim_{\epsilon \downarrow 0} 
R_{\rm act}(\epsilon) = R_*.$$
\item[$(c)$]
The Chernoff rate $R_*=R_*(\clW)$ 
satisfies:
$$\sup_{\alpha \in (1/2,1)} ( 1 - \alpha ) C_{\alpha}(\clW) 
\leq R_*(\clW) \leq \sup_{\alpha \in (0,1)}( 1 - \alpha ) 
C_{\alpha}(\clW).$$
\end{itemize}
\end{theorem}

\subsection{More general target rates}
\label{s:scassym}

As noted already earlier,
once the error probability is constrained 
to be below a given $\epsilon > 0$, 
the sample complexity 
formulation {\em fixes} the channel communication rate,
essentially 
mandating that optimal communication at blocklength 
$n^*_{\rm avg}\left(\mathcal{W},\epsilon \right)$ 
occurs at rate $R_\text{\rm act}(\epsilon)$ bits/transmission. 
Although this gives satisfying theoretical clarity,
the fact that it forces a specific rate may be quite 
restrictive in practice. Indeed, from the
practical point of view,
the symmetric constraint on the error probability
and the reciprocal of the codebook size is rather
artificial.

There are several ways to rectify this limitation.
First, a simple
approach is to examine the sample complexity when 
different weights 
are given to the two different objectives.
Specifically, we could define
the sample complexity in terms of
$$\pi P_{e, {\rm avg}}^{(n)} \quad \text{and} \quad (1-\pi)\frac{1}{|C_n|},$$
instead of the original 
$P_{e, {\rm avg}}^{(n)}$ and $1/|C_n|$,
for an appropriately chosen $\pi \in (0,1)$.

Using 
the random coding upper bound 
on $P_{e, {\rm avg}}^{(n)}$ combined with recent 
asymmetric hypothesis testing results~\cite{pensia:24} 
for the lower bound, it is straightforward to 
obtain corresponding upper and lower bounds to 
sample complexity in this case, as well as to evaluate 
the resulting actual communication rate.
This leads to useful results for moderate values of $\epsilon$,
but when $\epsilon$ is small the optimal blocklength 
given by the sample complexity necessarily grows, 
and the effects of the prior parameter $\pi$ 
get diminished. 

In the small-$\epsilon$ regime, and in fact for 
the entire range of $\epsilon\in(0,1)$, 
a more principled approach 
is to define, for any $\beta>0$, the {\em asymmetric 
channel coding sample complexity} of a channel $\clW$ as:
$$n^*_{\rm avg}(\mathcal{W},\epsilon;\beta) 
= \inf \left\{ n \geq 1 : \inf_{(C_n,d_n)} \max \Big\{ 
P^{(n)}_{e,{\rm avg}} , \frac{2^{n\beta}}{|C_n|} 
\Big\}\leq\epsilon \right\}.$$
This formulation clearly leads to much 
greater control of the target 
rate. In particular, it leads to an 
{\em asymmetric actual rate} given by
$$R_{\rm act}(\epsilon;\beta):= 
\beta + \frac{\log ( 1 / \epsilon )}
{n^*_{\rm avg}(\mathcal{W},\epsilon;\beta)}.$$

Getting accurate upper and lower bounds on 
$n^*_{\rm avg}(\mathcal{W},\epsilon;\beta)$
and evaluating
$R_{\rm act}(\epsilon;\beta)$ for small
$\epsilon$ are interesting 
avenues for further work.
We will not pursue these directions further here, 
except to note the following simple upper bound on 
$n^*_{\rm avg}\left(\mathcal{W},\epsilon;\beta\right)$, in the spirit
of the corresponding upper bound
on $n^*_{avg}\left(\mathcal{W},\epsilon\right)$ in 
Theorem~\ref{scDMC}.

\begin{theorem}[Asymmetric sample complexity bounds for DMCs]
\label{thm:asymmetric}
$\;$ For any DMC $\mathcal{W}$ with $C_{1/2}(\clW) > 0$, 
for all $\epsilon\in(0,1)$ and any $0<\beta<C(\clW)$
the asymmetric channel coding sample complexity 
satisfies,
$$
n^*_{\rm avg}(\mathcal{W},\epsilon;\beta) 
\leq \frac{\log(1/\epsilon)}{R(\clW,\beta)}
+ 1,
$$
where
$$R(\clW,\beta):=
\sup_{\alpha\in(1/2,1)}\big[( 1 - \alpha)( C_\alpha(\clW) - \beta )^+\big].
$$
\end{theorem}

\noindent
{\sc Proof.} Recalling from Proposition~\ref{prop:halfcap} that
$$\sup_P E_0(\rho,P) = \rho C_{\frac{1}{1+\rho}}(\clW),$$
the random coding upper bound in Proposition~\ref{prop:rcu} 
gives, for all $\rho \in (0,1)$,
\begin{align*}
\inf_{(C_n,d_n)} \max \left\{ P^{(n)}_{e,{\rm avg}}, 
\frac{2^{n\beta}}{|C_n|} \right\} 
&= 
	\inf_{0 < R \leq \log|A|}
	\;\;\inf_{(C_n,d_n)\;:\;|C_n| = \lceil 2^{nR}\rceil}
	\max \left\{ P^{(n)}_{e,{\rm avg}}, \frac{2^{n\beta}}{|C_n|}\right\}\\
&\leq 
	\inf_{0 < R \leq \log|A|} \max \left\{ 2^{-n \rho 
\big( C_{\frac{1}{1+\rho}}(\clW) - R \big)} , 
2^{-n \left( R - \beta \right)} \right\} \\
    &\leq 2^{-n \left( 1 - \alpha \right) \left( C_\alpha(\clW) - \beta 
	\right)},
\end{align*}
where $\alpha = \frac{1}{1+\rho}$ ranges from $1/2$ to $1$, and where 
the last line is obtained by choosing $R = (1-\alpha)C_\alpha(\clW) 
+ \alpha \beta$. The claimed result follows by the definition
of $n^*_{\rm avg}(\clW,\epsilon;\beta)$.
\qed

We close this section with two simple observations. First, 
by the definition of the $C_\alpha(\clW)$ and the assumption 
that $0<\beta<C(\clW)$, there is always an interval
of values $\alpha$ in $(1/2,1)$ for which $C_\alpha(\clW)-\beta$ is 
strictly positive, so the upper bound in the theorem is not vacuous. 
Second, since by definition we always have that
$n^*_{\rm avg}(\mathcal{W},\epsilon) 
\leq n^*_{\rm avg}(\mathcal{W},\epsilon;\beta)$, 
by Theorem~\ref{scDMC} we also always have the elementary 
lower bound
$$n^*_{\rm avg}(\mathcal{W},\epsilon;\beta)
\geq\frac{\log(1/\epsilon)-2}{C_{1/2}(\clW)}.$$

\section{The additive white Gaussian noise channel}
\label{sec:scgauss}

The AWGN channel with noise power $\sigma^2>0$
is a memoryless channel 
$\clW=\{W^n\}$ from $A=\RL$ to $B=\RL$ such that
the conditional distribution $W(\cdot|x)$
of the output $Y$ given an input $X=x$ is
$N(x,\sigma^2)$.
We impose a quadratic power constraint
on the input: Given $P>0$ we assume
that the codebook $C_n$ 
of any code $(C_n,d_n)$
either satisfies the power constraint
for every codeword in $C_n$ individually (an ``almost sure'' 
power constraint),
or on the average (an ``expected power'' constraint).

Let $\clC_{\rm a.s.}(n,P)$ be the collection
of all codebooks $C_n$
of blocklength $n$
such that $\|x^n\|^2\leq nP$ for all $x^n\in C_n$,
where $\|\cdot\|$ denotes the Euclidean norm on $\RL^n$.
Similarly let
$\clC_{\rm exp}(n,P)$ be the corresponding collection
of all codebooks $C_n$ of blocklength $n$
such that 
$$\frac{1}{|C_n|}\sum_{x^n\in C_n}\|x^n\|^2\leq nP,$$

The {\em channel coding sample complexity} of the AWGN channel
with respect to the almost sure power constraint is then defined as,
$$n^*_{\rm AWGN,a.s.}(\sigma^2,P,\epsilon) 
= \inf \left\{ n \geq 1 : \inf_{(C_n,d_n)\;:\;C_n\in\clC_{\rm a.s.}(n,P)}
\max \Big\{ P^{(n)}_{e,{\rm avg}} , \frac{1}{|C_n|}\Big\} 
\leq \epsilon \right\},$$
and the channel coding sample complexity of the AWGN channel
with respect to the expected power constraint is similarly defined as,
$$n^*_{\rm AWGN,exp}(\sigma^2,P,\epsilon) 
= \inf \left\{ n \geq 1 : 
\inf_{(C_n,d_n)\;:\;C_n\in\clC_{\rm exp}(n,P)}
\max \Big\{ P^{(n)}_{e,{\rm avg}} , \frac{1}{|C_n|}\Big\} 
\leq \epsilon \right\}.$$

Note that, trivially, we always have
$\clC_{\rm a.s.}(n,P)\subset\clC_{\rm exp}(n,P)$,
so that
\be
n^*_{\rm AWGN,exp}(\sigma^2,P,\epsilon) 
\leq n^*_{\rm AWGN,a.s.}(\sigma^2,P,\epsilon).
\label{eq:asexp}
\ee

\noindent
{\bf N.B.} 
Throughout this section we use the natural logarithm 
$\log_e$, and all rates and information-theoretic functionals
are expressed in nats.

\subsection{Preliminaries}
\label{s:contiprelim}

Let $P,Q$ be two probability measures on the
same measurable space $(A,\clA)$ and let
$\lambda$ be a dominating $\sigma$-finite measure
on $(A,\clA)$, i.e., such that 
$P$ and $Q$ are both absolutely continuous
with respect to $\lambda$. We write
$p,q$ for their corresponding densities with
respect to $\lambda$.
The {\em R\'{e}nyi divergence of order $\alpha\in(0,1)$}
between $P$ and $Q$ is defined as
$$D_{\alpha}(P\|Q):=\frac{1}{\alpha-1}\log_e\left(\int p^\alpha q^{1-\alpha}
\,d\lambda
\right).
$$

For any pair of random variables $X,Y$ taking values
in the general spaces $A$ and $B$, respectively,
we can proceed exactly as in Section~\ref{s:Dalpha}
to define the Sibson and 
Augustin--Csisz\'{a}r mutual informations
by replacing
the relative entropy by the R\'{e}nyi divergence.
Similarly,
for any channel $W=(W_{Y|X})$ between $A$ and $B$, 
with or without
input power constraints, we can define
the corresponding 
Sibson and 
Augustin--Csisz\'{a}r capacities
by taking the corresponding suprema
over all input distributions,
possibly satisfying any required
power constraints.

In the case of general power-constrained channels,
the Sibson and 
Augustin--Csisz\'{a}r capacities need not
be equal, 
therefore, for the AWGN channel 
with noise power $\sigma^2$ and power constraint $P$,
we concentrate
on the 
Augustin--Csisz\'{a}r capacity of order $\alpha\in(0,1)$,
defined as,
$$
C_{\alpha}^{\rm AC}(\sigma^2,P):=\sup_{X\;:\;\BBE(X^2)\leq P} \inf_{Q}
\BBE\big[D_\alpha\big(W(\cdot|X)\|Q\big)\big],$$
where the supremum is over all real-valued random variables
$X$ with $\BBE(X^2)\leq P$, the infimum
is over all probability measures $Q$ on $\RL$,
and $W(\cdot|x)$ is the $N(x,\sigma^2)$ distribution.

In the case of the AWGN channel with parameters
$\sigma^2$ and $P$, following~\cite{gallager:book}
we define,
for any $\rho, r \geq 0$, and any density $g$ of $\RL$,
$$E_0(\rho,g,r) := - \log_e \left( \int_{\mathbb{R}} \left( \int_{\mathbb{R}} 
e^{r \left( x^2 - P \right)} g(x) w(y|x)^{\frac{1}{1+\rho}}dx 
\right)^{1+\rho} dy\right),
$$
where $w(\cdot|x)$ denotes the $N(x,\sigma^2)$ density.
Taking $g$ to be the $N(0,P)$ density $\phi_P$, we can then
define the following natural analog of 
Gallager's $E_0$ function we saw earlier, via,
\begin{align}
E_0(\rho,\gamma) 
&:= 
	\sup_{r \geq 0} 
	E_0(\rho,\phi_P,r) \nonumber \\
&= 
	\sup_{0 \leq \lambda < \frac{1}{2\gamma}}
	\left[ (1+\rho)\lambda\gamma + \frac{1}{2} 
	\log_e \left(1 - 2\lambda\gamma \right) + \frac{\rho}{2} 
	\log_e \left( 1 - 2\lambda\gamma + \frac{\gamma}{1+\rho} \right)
	\right],
	\label{eq:gallagerexpgauss}
\end{align}
where $\gamma:=P/\sigma^2$ denotes the SNR
and the second expression follows by the substitution 
$\lambda=r\sigma^2$.
The corresponding random coding exponent is
\be
E_r(R,\gamma) := \max_{\rho \in [0,1]}\big[E_0(\rho,\gamma) 
- \rho R\big] = E_0(\rho^*,\gamma) - \rho^* R,
\label{eq:rcexpAWGN}
\ee
where $\rho^* = \rho^*(R,\gamma)$ denotes a maximizer. 

The following 
proposition directly connects the Gallager function to the 
Augustin--Csisz\'{a}r capacity of order $1/2$.
Its proof, mostly based on~\cite{verdu:21},
is given in Appendix~\ref{app:capacity}.

\begin{proposition}[AWGN channel Augustin--Csisz\'{a}r capacity]
\label{prop:AWGNcutoff}
$\;$ The Augustin--Csisz\'{a}r capacity of order~1/2 
of the AWGN channel with
parameters $\sigma^2$ and $P$ depends only on the SNR 
$\gamma=P/\sigma^2$
and is given by
\be
C_{1/2}^{\rm AC}(\sigma^2,P)
=
C_{1/2}(\gamma):=
\frac{1}{2}\left( 1 + \frac{\gamma - \Delta}{2} \right) 
+ \frac{1}{2}\log_e \left( \frac{1}{2} + \frac{\Delta}{4} \right),
\label{eq:Gcapacity}
\ee
where $\Delta = \sqrt{4 + \gamma^2}$.
Moreover, $C_{1/2}(\gamma)$ can be expressed as,
\be
C_{1/2}(\gamma) 
= E_0(1,\gamma) = E_0(1,\phi_P,r^*),
\label{eq:Gcapacity2}
\ee
where $\phi_P$ denotes the $N(0,P)$ density and
\be
0<r^* := \frac{1}{4P} \left( 1 + \frac{\gamma - \Delta}{2} \right)
<\frac{1}{2P}.
\label{eq:Gcapacity3}
\ee
\end{proposition}

\subsection{New error probability bounds}

The best achievable error probability for the AWGN channel has been studied 
extensively since Shannon's 1959 work~\cite{shannon:59sp}.
Gallager's random coding upper bound extends to the
AWGN channel~\cite{gallager:book}, but with a prefactor that 
grows polynomially with the blocklength $n$.
The random-coding union (RCU) bound of Polyanskiy, Poor and 
Verd\'u~\cite{PPV:10} applies directly to Shannon's spherical code ensemble. 
Erseghe~\cite{erseghe:16} gives integral representations and numerical 
methods for evaluating this bound, and M\"uller~\cite{muller:21} revisits 
the exact ensemble-average error probability and its computation. Tan and 
Tomamichel~\cite{tan:15} determine the third-order term in the normal 
asymptotic approximation for the AWGN channel, and they also discuss 
the classical error asymptotics, whose prefactor decreases polynomially 
with the blocklength.

In particular, at any fixed SNR and at rates above the critical rate 
and below capacity, these asymptotics already imply an upper 
bound with some blocklength-independent prefactor. But this 
approach necessarily introduces an unspecified multiplicative constant 
factor in the bounds.

In this section we develop explicitly computable, nonasymptotic upper 
bounds for the error probability, that maintain the random-coding 
exponent and only include a prefactor independent of the 
blocklength.

Let $\epsilon^*(n,R,\sigma^2,P)$ denote the best achievable
average error probability for the AWGN channel with 
noise power $\sigma^2$ and a quadratic almost-sure
power constraint $P$:
$$ 
\epsilon^*(n,R,\sigma^2,P)=\inf\left\{P_{e,{\rm avg}}^{(n)}:
\mbox{codes}\;(C_n,d_n)\;\mbox{with}\;C_n\in\clC_{\rm a.s.}(n,P),
\;|C_n|\geq e^{nR}
\right\}.$$

Our first bound is based on a refined analysis 
of Shannon's spherical code construction,
and applies at every rate below the AWGN channel capacity
$$C(\gamma):=\frac{1}{2}\log_e(1+\gamma).$$
Its proof
is given in Appendix~\ref{app:probAWGN1}.
Its exponent is the classical random coding
exponent $E_r(R,\gamma)$ and the constant prefactor
is an explicit function of the rate $R$ and
the SNR $\gamma$.

\begin{theorem}
\label{thm:probAWGN1}
Consider the AWGN channel with noise power $\sigma^2>0$, input power
constraint $P>0$, and SNR $\gamma=P/\sigma^2$. For any rate
$0<R<C(\gamma)$ and every $n\geq 1$,
\bqq
\epsilon^*(n,R,\sigma^2,P)\leq D(R,\gamma)e^{-n E_r(R,\gamma)},
\label{eq:probAWGN1}
\eqq
where, 
$$
D(R,\gamma):=1+2^{1-\rho^*}e^{E_0(\rho^*,\gamma)},
$$
with $\rho^*$ as in~\eqref{eq:rcexpAWGN}.
In particular,
\bqq
\epsilon^*(n,R,\sigma^2,P)\leq[1+2\sqrt{1+\gamma}]e^{-n E_r(R,\gamma)}.
\label{eq:probAWGN1snrconstant}
\eqq
\end{theorem}

The prefactor in~\eqref{eq:probAWGN1}
can also be chosen as a function of the 
rate only.
The proof of Theorem~\ref{thm:probAWGN2}, which requires a second 
coding construction at high SNR,
is given in Appendix~\ref{app:probAWGN2}.

\begin{theorem}
\label{thm:probAWGN2}
For each $R>0$, let
\begin{align*}
\gamma_0(R)&:=16(1+R)e^{2R},\\
B(R)&:=1+2\sqrt{1+\gamma_0(R)}.
\end{align*}
For every $n\geq 1$ and all $\gamma>e^{2R}-1$,
\bqq
\epsilon^*(n,R,\sigma^2,P)\leq B(R)e^{-n E_r(R,\gamma)}.
\label{eq:probAWGN2}
\eqq
If $\gamma\geq \gamma_0(R)$, the stronger estimate holds:
\bqq
\epsilon^*(n,R,\sigma^2,P)\leq e^{-n E_r(R,\gamma)}.
\label{eq:probAWGN2bis}
\eqq
\end{theorem}

In Appendix~\ref{app:probAWGN3} we show that,
combining Theorems~\ref{thm:probAWGN1} and~\ref{thm:probAWGN2},
yields the following convenient error probability
bound at the specific rate $R =\frac{1}{2} C_{1/2}(\gamma)$.

\begin{corollary}
\label{cor:probAWGN3}
For every $n\geq 1$,
\bqq
\epsilon^*\Big(n,\frac{1}{2}C_{1/2}(\gamma),\sigma^2,P\Big)
\leq \left(1+e^{C_{1/2}(\gamma)}\right) e^{-\frac{n}{2} C_{1/2}(\gamma)}.
\label{eq:probAWGN3.1}
\eqq
Moreover, if $\gamma \geq 1300$,
\bqq
\epsilon^*\Big(n,\frac{1}{2}C_{1/2}(\gamma),\sigma^2,P\Big)
\leq e^{-\frac{n}{2} C_{1/2}(\gamma)}.
\label{eq:probAWGN3.2}
\eqq
\end{corollary}

\subsection{Sample complexity of the AWGN channel}
\label{s:awgnnstar}

We are now ready to state and prove our 
sample complexity bounds for the AWGN channel.

\begin{theorem}[Sample complexity of the AWGN channel]
\label{scGauss}
Consider the AWGN channel with noise power $\sigma^2$, 
input power constraint $P$, SNR $\gamma = P/\sigma^2$,
and Augustin--Csisz\'{a}r cost-constrained capacity $C_{1/2}(\gamma)>0$.
For all $\sigma^2>0$, $P>0$ and all $\epsilon \in (0,1)$, we have:
\begin{align}
\frac{\log_e (1/\epsilon) - 2}{C_{1/2}(\gamma)} 
&\leq 
	n^*_{\rm AWGN,exp}(\sigma^2,P,\epsilon)
	\label{eq:awgnexp}\\
&\leq
	n^*_{\rm AWGN,a.s.}(\sigma^2,P,\epsilon)	
\leq 
	\frac{2 \big( \log_e(1/\epsilon) + 1 \big)}{C_{1/2}(\gamma)} + 3.
	\label{eq:awgnaslowSNR}
\end{align}
Moreover, for $\gamma\geq 1300$, the upper bound can be tightened to:
\bqq
n^*_{\rm AWGN,a.s.}(\sigma^2,P,\epsilon) 
\leq \frac{2 \log_e (1/\epsilon)}{C_{1/2}(\gamma)} + 1.
\label{eq:awgnashighSNR}
\eqq
\end{theorem}

\noindent
{\sc Proof.}
The proof of the lower bound in~\eqref{eq:awgnexp} is essentially
identical to the proof of the corresponding bound
in Theorem~\ref{scDMC} in the discrete case,
where we again use the Le Cam lower bound 
in Theorem~\ref{thm:lecam} in combination with the 
general-alphabet version of Proposition~\ref{prop:DTV}
in~\cite[Lemma~2.3]{tsybakov:09}. The only additional
step is to note that $\hat{X}\sim\hat{P}$ will satisfy
$\BBE[\hat{X}^2]\leq P$ for any codebook $C_n$
under consideration, so that the final infimum
can be taken over all input distributions $P$
such that $\BBE[X^2]\leq P$ when $X\sim P$.

The first inequality in~\eqref{eq:awgnaslowSNR}
is simply~\eqref{eq:asexp}.
For the upper bound in~\eqref{eq:awgnaslowSNR}
we can apply the bound~\eqref{eq:probAWGN3.1}
in Corollary~\ref{cor:probAWGN3} to obtain that,
\begin{align*}
& 
	\inf_{(C_n,d_n)\;:\;C_n\in\clC_{\rm a.s.}(n,P)}
	\max \Big\{ P^{(n)}_{e,{\rm avg}} , \frac{1}{|C_n|}\Big\} \\
&= 
	\inf_{R>0}\; 
	\inf_{C_n\in\clC_{\rm a.s.}(n,P),\,|C_n|=\lceil e^{nR}\rceil}
	\max \Big\{ P^{(n)}_{e,{\rm avg}} , 
	\frac{1}{\lceil e^{nR} \rceil} \Big\} \\
&\leq 
	\inf_{R>0}\max 
	\big\{\epsilon^*\left(n,R,\sigma^2,P\right), e^{-nR}\big\} \\
&\leq 
	\max \Big\{\epsilon^*\Big(n,\frac{1}{2}C_{1/2}(\gamma),
	\sigma^2,P\Big), e^{-\frac{n}{2}C_{1/2}(\gamma)}\Big\} \\
&\leq 
	\big(1+e^{C_{1/2}(\gamma)}\big)e^{-\frac{n}{2}C_{1/2}(\gamma)},
\end{align*}
which, combined with the definition 
of $n^*_{\rm AWGN,a.s.}(\sigma^2,P,\epsilon)$, implies that
\be
n^*_{\rm AWGN,a.s.}(\sigma^2,P,\epsilon)
\leq
\frac{2 \big[ \log_e(1/\epsilon) + 
\log_e \big( 1+e^{C_{1/2}(\gamma)}\big)\big]}{C_{1/2}(\gamma)} + 1.
\label{eq:medSNR}
\ee
The required inequality 
in~\eqref{eq:awgnaslowSNR} follows from~\eqref{eq:medSNR}
via an application of the elementary inequality
$\log_e(1+e^x)\leq 1+x$, $x\geq 0$.

The proof of the corresponding inequality~\eqref{eq:awgnashighSNR} 
in the regime where $\gamma\geq 1300$ is exactly the same, simply 
using the bound~\eqref{eq:probAWGN3.2}
in Corollary~\ref{cor:probAWGN3} instead of~\eqref{eq:probAWGN3.1}.
\qed

The same remarks as those made in the case of DMCs
after Theorem~\ref{scDMC} apply here as well.
In particular, the actual communication rate
$R_{\rm act}(\epsilon)$ can be defined as before,
and its behavior for $\epsilon$ close to zero
would be interesting to investigate in connection
with the AWGN channel's reliability function.
Also, more general target rates may be considered
by adapting the definition of $n^*_{\rm avg}(\clW,\epsilon;\beta)$
to the present setting.
Numerical comparisons of the bounds
in Theorem~\ref{scGauss} with the
corresponding estimates to the sample complexity
$n^*_{\rm AWGN,a.s.}(\sigma^2,P,\epsilon)$
obtained via the Gaussian approximation in~\cite{PPV:10}
reveal qualitatively similar results to those displayed
in Section~\ref{s:nonasympvsgaus}
in the case of DMCs.

Finally, we note that the stronger upper bound~\eqref{eq:medSNR}
was actually established in the proof. It is easy to find
examples when $C_{1/2}(\gamma)$ is relatively small,
equivalently, for small enough SNRs $\gamma>0$, where
the bound~\eqref{eq:medSNR} is significantly more accurate
than~\eqref{eq:awgnaslowSNR}
in the theorem.

\section*{Acknowledgments}

We wish to thank Albert Guill\'{e}n i F\`{a}bregas 
for his useful comments on this work.
ChatGPT-6 was used in the analysis of the numerical
examples in Section~\ref{s:nonasympvsgaus}
and for minor editorial help with the exposition.
The mathematical results and their proofs were
developed and written by the authors.

\begin{appendices}

\section{Proof of Theorem~\ref{thm:limitingactualrate}}
\label{app:actual}

Part~$(a)$. Write
$$e_n(R):=-\frac1n\log\epsilon^*(n,R).$$
First, suppose there is an $\eta>0$ and 
a sequence $\{\epsilon_k\}$ decreasing to zero,
such that
$$R_{\rm act}(\epsilon_k)\ge R_*+\eta.$$
Set $n_k=n^*_{\mathrm{avg}}(\mathcal W,\epsilon_k)$. A feasible code 
at blocklength $n_k$ has rate at least $R_*+\eta\geq R_*$, 
hence,
$$
e_{n_k}(R_*)
\geq
\frac{\log(1/\epsilon_k)}{n_k}
\geq R_*+\eta.
$$
But, since
$$ n_k\geq \frac{\log(1/\epsilon_k)}{\log|A|}
\to\infty,$$
this contradicts
$$
\limsup_{n\to\infty}e_n(R_*)=E^*(R_*)=R_*,
$$
proving that 
$$
\limsup_{\epsilon\to 0} R_{\rm act}(\epsilon)\leq R_*.
$$

To establish the reverse limsup inequality,
fix $0<\eta<R_*$ and note that, by the definitions,
for infinitely many $n$ we have
$$\epsilon^*(n,R_*)\le 2^{-n(R_*-\eta)}.$$
For those $n$, put $\epsilon_n=2^{-n(R_*-\eta)}$. 
The corresponding codes have both error probability at most 
$\epsilon_n$ and size at least $1/\epsilon_n$. Thus,
$n^*_{\mathrm{avg}}(\mathcal W,\epsilon_n)\le n$,
and 
$R_{\rm act}(\epsilon_n)\ge R_*-\eta$. Since $\eta$ was arbitrary,
this establishes that,
$$
\limsup_{\epsilon\to 0} R_{\rm act}(\epsilon)\geq R_*,
$$
completing the proof of part~$(a)$.

Part~$(b)$. Here we need to establish the corresponding
lower bound to~\eqref{eq:parta},
for the $\liminf$ as ${\epsilon\downarrow 0}$.
Choose and fix
$0<\eta <R_*$ arbitrary.
By the assumption of part~$(b)$,
for all sufficiently large $n$ depending on $\eta$,
\be
\epsilon^*(n,R_*) \leq 2^{-n [ E^*(R_*) - \eta] }. 
\label{scmaxavgproof1}
\ee
In particular, letting
$$n_\epsilon := \left\lceil \frac{\log ( 1/\epsilon)}
{R_* - \eta} \right\rceil,
$$
the bound in~\eqref{scmaxavgproof1} holds 
with $n_\epsilon$ in place of $n$,
for all $\epsilon > 0$ small enough depending again on $\eta$.
Also, by the definition
of $n_\epsilon$ we have that 
both $n_\epsilon\left[E^*(R^*)-\eta\right] \geq \log \left( 1/\epsilon \right)$ 
and $n_\epsilon R_* \geq \log \left( 1/\epsilon \right)$. Hence 
for $\epsilon > 0$ small enough we have that both,
$$\epsilon^*(n_\epsilon,R_*) \leq \epsilon \quad \text{and} 
\quad \frac{1}{\lceil 2^{n_\epsilon R_*} \rceil} \leq \epsilon,$$
which 
implies that $n^*_{\rm avg}(\mathcal{W},\epsilon) 
\leq n_\epsilon$. Therefore,
$$\liminf_{\epsilon \rightarrow 0} \frac{\log (1/ \epsilon )}
{n^*_{\rm avg}(\mathcal{W},\epsilon)} 
\geq \liminf_{\epsilon \rightarrow 0} \frac{\log(1/\epsilon)}{n_\epsilon} 
\geq R_* - \eta.$$
Since $0<\eta<R_*$ was arbitrary,
we obtain,
$$
\liminf_{\epsilon \to 0} R_{\rm act}(\epsilon) \geq R_*,
$$
completing the proof of part~$(b)$.

Part~$(c)$. 
Let
$$
G(\rho):=\sup_P E_0(\rho,P).$$
Then the random-coding upper bound in Proposition~\ref{prop:rcu}
applied to $R_*$ gives
$G(\rho)-\rho R_*\le R_*$
for every $0<\rho<1$,
and hence,
$$
R_*\ge \sup_{0<\rho<1}\frac{G(\rho)}{1+\rho}.$$
Reparametrizing in terms of $\alpha=\frac{1}{1+\rho}$
and applying Proposition~\ref{prop:halfcap}~$(iii)$
gives the lower bound in part~$(c)$.

For the upper bound, write
$$
H:=\sup_{\rho>0}\frac{G(\rho)}{1+\rho}.
$$
If $R_*>H$, then for every $\rho>0$,
$G(\rho)-\rho R_* \leq H-\rho(R_*-H) \leq H$.
The sphere-packing bound~\cite{gallager:book} would 
therefore imply
$$ R_*=E^*(R_*)\leq E_{\rm sp}(R_*)\leq H<R_*,$$
a contradiction. Thus 
$$R_*\le H=\sup_{\rho>0} \frac{G(\rho)}{1+\rho}
=
\sup_{\rho>0} \frac{\rho}{1+\rho}C_{1/(1+\rho)}(\mathcal W)
=
\sup_{\alpha\in(0,1)}(1-\alpha)C_\alpha(\mathcal W),
$$
as required.
\qed

\section{Proof of Proposition~\ref{prop:AWGNcutoff}}
\label{app:capacity}

The expression for $C_{1/2}^{\rm AC}(\sigma^2,P)$ 
in~\eqref{eq:Gcapacity} and the first equality
in~\eqref{eq:Gcapacity2} were 
both established
in~\cite[Theorem~23]{verdu:21}. 

To establish that $E_0(1,\gamma)=E_0(1,\phi_P,r^*)$ with
$r^*$ as in~\eqref{eq:Gcapacity3},
first recall from~\cite[Eq.~(7.4.21)]{gallager:book}
that 
$$E_0 (1,\phi_P , r ) = 2rP + \frac{1}{2} \log_e (1 - 2rP ) 
+ \frac{1}{2} \log_e \left( 1 - 2rP + \frac{P}{2\sigma^2} \right),$$
for all $0 \leq r < \frac{1}{2P}$,
while
$E_0(1, \phi_P, r) = -\infty$ for $r \geq \frac{1}{2P}$. 
Differentiating,
$$\frac{\partial}{\partial r} E_0 \left(1,\phi_P, r \right) 
= 2P - \frac{P}{1 - 2rP} - \frac{P}{1 - 2rP + \frac{P}{2\sigma^2}},$$
so $E_0 (1,\phi_P, r )$ is nondecreasing on $(0,r^*)$ 
and nonincreasing on $(r^*,\frac{1}{2P})$, where $r^*$ 
is the unique zero of the above derivative
smaller than $\frac{1}{2P}$. 
Solving $\frac{\partial}{\partial r}E_0(1,\phi_P,r)=0$
for $r$
yields the value of $r^*$ in~\eqref{eq:Gcapacity3}
as claimed.
\qed

\section{AWGN channel error probability bounds}

We first record some elementary properties 
of the $E_0$ function defined in~\eqref{eq:gallagerexpgauss} 
and of the random coding error exponent $E_r$ defined in~\eqref{eq:rcexpAWGN}.
For $0<\rho\leq 1$, let $\beta_\rho$ be the unique solution
in $(0,1)$ of
\bqq
 t_\rho(\beta):=\frac{\rho}{\beta}-\frac{1}{1-\beta}-\gamma=0.
\label{eq:beta}
\eqq
Define, 
\begin{align*}
 G_{\rho,\gamma}(\beta)&:=\rho\log_e\beta+\log_e(1-\beta)-\gamma\beta,\\
 g(\rho)&:=(1+\rho)\log_e(1+\rho)-\rho\log_e\rho,
\end{align*}
for $0<\beta<1$ and $\rho>0$, 
respectively.

\begin{proposition}[Exponent properties]
\label{prop:exponent}
For $0<\rho\leq 1$ and $\gamma>0$, we have:
\begin{align}
 E_0(\rho,\gamma)&=-\tfrac12[G_{\rho,\gamma}(\beta_\rho)+g(\rho)],
                                                        \label{eq:E0alt}\\
 \frac{\partial}{\partial\rho} E_0(\rho,\gamma)&=\tfrac{1}{2}\log_e
                  \Big(\frac{\rho}{(1+\rho)\beta_\rho}\Big),
						\label{eq:derivative}\\
 0\leq E_0(\rho,\gamma)&\leq
       \min\left\{\rho C(\gamma),\;\frac{\gamma\rho}{2(1+\rho)}\right\}.
                                                        \label{eq:E0bounds}
\end{align}
Moreover, $E_0(0,\gamma)=0$, 
$$\frac{\partial}{\partial \rho} E_0(0+,\gamma)=C(\gamma),$$
and, for $0<R<C(\gamma)$, every maximizer in~\eqref{eq:rcexpAWGN} is strictly
positive and
\bqq
\label{eq:Erbounds}
0<E_r(R,\gamma)\leq C(\gamma)-R.
\eqq
\end{proposition}

\noindent
{\sc Proof.}
The function $t_\rho$ is strictly decreasing, with limits $+\infty$ and
$-\infty$ at the two endpoints, so $\beta_\rho$ exists and is unique.
The definition of $\beta_\rho$ 
in~\eqref{eq:beta} gives
\bqq
\begin{split}
0<\beta_\rho&<\frac{\rho}{1+\rho},\\
1+\rho-\gamma&\beta_\rho=\frac{1}{1-\beta_\rho}.
\end{split}
\label{eq:betabounds}
\eqq
The expression being maximized in~\eqref{eq:gallagerexpgauss} is strictly 
concave in $\lambda$.
Substituting $\lambda^*=\beta_\rho/[2(1+\rho)]$ in its derivative
and using~\eqref{eq:beta} and~\eqref{eq:betabounds}, shows that this is its
unique maximizer and yields~\eqref{eq:E0alt}. This is a change of
variables in Gallager's Gaussian 
calculation~\cite[Equations~(124)--(128)]{gallager:65}; his parameter $\beta$, 
denoted here
by $b$, satisfies $b=1-\gamma\beta_\rho/(1+\rho)$.

Since 
$$
\frac{\partial t_\rho(\beta)}{\partial\beta}
=-\frac{\rho}{\beta^2}-\frac{1}{(1-\beta)^2}<0,
$$
the implicit function theorem implies that $\beta_\rho$ is continuously 
differentiable in $\rho$, with $\gamma$ fixed. 
Differentiating~\eqref{eq:E0alt} and using the fact that
$\frac{\partial}{\partial\beta} G_{\rho,\gamma}(\beta_\rho)=0$, 
therefore, 
gives~\eqref{eq:derivative}. Also,
\bqq
\begin{split}
\frac{\rho}{\beta_\rho}=\gamma&+\frac{1}{1-\beta_\rho},\\
\lim_{\rho\downarrow 0}
\frac{\beta_\rho}{\rho} & = \frac{1}{1+\gamma}.
\end{split}
\label{eq:betaratio}
\eqq
At $\rho=0$, the maximum in~\eqref{eq:gallagerexpgauss} 
is achieved at $\lambda=0$ 
and equals zero, and the expression in~\eqref{eq:E0alt} 
admits the same extension by continuity. Therefore,
$$\frac{\partial}{\partial\rho} E_0(0+,\gamma)=C(\gamma),$$
and by~\eqref{eq:betabounds} and~\eqref{eq:betaratio},
\begin{align*}
 \partial_\rho E_0(\rho,\gamma)
 &\leq\frac{1}{2}\log_e\Big(\frac{\gamma+1+\rho}{1+\rho}\Big)\\
 &\leq C(\gamma).
\end{align*}
Integrating proves the first upper bound in~\eqref{eq:E0bounds}.
Since $G_{\rho,\gamma}$ is maximized at $\beta_\rho$, comparison with
$\beta=\rho/(1+\rho)$ gives the second. Nonnegativity follows by taking
$\lambda=0$ in~\eqref{eq:gallagerexpgauss}. Finally,
$E_0(\rho,\gamma)-\rho R$ is continuous on $[0,1]$ and has a positive
right derivative at zero. Its maximum is positive and attained away from
zero, while~\eqref{eq:E0bounds} implies~\eqref{eq:Erbounds}.
\qed

\subsection{Proof of Theorem~\ref{thm:probAWGN1}}
\label{app:probAWGN1}

We employ Shannon's~\cite{shannon:59sp} codebook construction. For 
$n\geq 2$,
consider a random codebook $C_n$ with $M=\lceil e^{nR}\rceil$
codewords $X^n(i)$, each drawn independently and identically distributed
to be uniform on the sphere of radius $\sqrt{nP}$ in $\RL^n$, so that
they satisfy the power constraint individually with equality. 
Note that the
codewords are distinct almost surely,
and equip $C_n$ with the maximum likelihood decoder.
Let $P_e(C_n)$ denote the average error probability
$P_{e,{\rm avg}}^{(n)}$ achieved by $C_n$.
To evaluate $\BBE[P_e(C_n)]$,
it suffices to consider the error probability when $X^n(1)$ is
transmitted, and by rotational symmetry we can assume that
$X^n(1)=\sqrt{nP}\,e_1$, where $e_1$ denotes the first
coordinate unit vector in $\RL^n$.

Let $\Theta\in[0,\pi]$ be the angle between $X^n(1)$ 
and the received sequence $Y^n$.
Conditional on these two vectors, another codeword leads to a pairwise
decoding error precisely when it belongs to the spherical cap of
half-angle $\Theta$ centered in the direction of $Y^n$ (ties have
zero probability). For a fixed angle $\theta$, the normalized 
area of this
cap is
$$
a_n(\theta)=c_n\int_0^\theta(\sin t)^{n-2}\,dt,
$$
where
$$
c_n=\frac{\Gamma(n/2)}{\sqrt\pi\,\Gamma((n-1)/2)}.
$$
Writing $a=a_n(\Theta)$, the other $M-1$ competing codewords remain
conditionally independent, and each causes a decoding error with
probability $a$. The conditional ensemble error probability is therefore
\begin{equation*}
 1-(1-a)^{M-1}\leq\min\{1,(M-1)a\}\leq[(M-1)a]^\rho,
\end{equation*}
for any $0<\rho\leq 1$. The expectation of the 
middle expression is the spherical RCU 
bound~\cite[Theorem~16]{PPV:10}. 

Since $M-1<e^{nR}$,
\bqq
\BBE[P_e(C_n)]\leq e^{n\rho R}\BBE[a_n(\Theta)^\rho].
\label{eq:ensemble}
\eqq
The standard cap estimate in~\cite[Proposition~5.1]{aubrun:book} states that
$$
 a_n(\theta)\leq\tfrac12(\sin\theta)^{n-1},\qquad
 0\leq\theta\leq \pi/2,
$$
for every $n\geq 2$. Splitting~\eqref{eq:ensemble} at $\Theta=\pi/2$
gives
\bqq
 \BBE[P_e(C_n)]\leq e^{n\rho R}
 \Big\{2^{-\rho}\BBE[(\sin\Theta)^{\rho(n-1)}]
                    +\BBP\Big(\Theta>\frac{\pi}{2}\Big)\Big\}.
\label{eq:split}
\eqq

As $Y^n/\sigma=\sqrt{n\gamma}\,e_1+Z^n$, defining
\begin{equation*}
V=\sqrt{n\gamma}+Z_1\quad\mbox{and}\quad U=\sum_{k=2}^n Z_k^2,
\end{equation*}
we have that $V\sim N(\sqrt{n\gamma},1)$ and $U\sim\chi^2_{n-1}$
are independent, and
\begin{equation*}
 \cos\Theta=\frac {V}{\sqrt{U+V^2}},\qquad
 \sin^2\Theta=\frac {U}{U+V^2}.
\end{equation*}
Now, the event $\{\Theta>\pi/2\}$ is exactly $\{V<0\}$, 
so a standard Gaussian
tail bound gives 
$$\BBP(\Theta>\pi/2)\leq e^{-n\gamma/2}.$$
Since the inequality~\eqref{eq:E0bounds} implies 
$E_0(\rho,\gamma)\leq\gamma/4$, from~\eqref{eq:split} we get
\bqq
\BBE[P_e(C_n)]\leq e^{n\rho R}
                   [2^{-\rho}J_{n,\rho}+e^{-nE_0(\rho,\gamma)}],
\label{eq:Jboundstart}
\eqq
where, with $d=(n-1)/2$,
$$
J_{n,\rho}:=\BBE[(\sin\Theta)^{\rho(n-1)}]
            =\BBE\left[\left(\frac {U}{U+V^2}\right)^{\rho d}\right].
$$

Most of the remainder of the proof
is devoted to the derivation of the 
following upper bound on $J_{n,\rho}$:

\begin{lemma}
\label{lem:moment}
For all $n\geq 2$, $0<\rho\leq 1$ and $\gamma>0$,
$$
J_{n,\rho}\leq 2e^{-(n-1)E_0(\rho,\gamma)}.
$$
\end{lemma}

\noindent
{\sc Proof.}
Let $\alpha=\rho d>0$. Using the Gamma integral for
$(U+V^2)^{-\alpha}$, Fubini's theorem and independence,
\bqq
J_{n,\rho}=\frac{1}{\Gamma(\alpha)}\int_0^\infty
t^{\alpha-1}\BBE[U^\alpha e^{-tU}]\BBE[e^{-tV^2}]\,dt.
\label{eq:laplace}
\eqq
Direct integration gives
\begin{align*}
 \BBE[U^\alpha e^{-tU}]&=
 2^\alpha\frac{\Gamma(d+\alpha)}{\Gamma(d)}(1+2t)^{-d-\alpha},\\
\mbox{and}\quad \BBE[e^{-tV^2}]&=(1+2t)^{-1/2}
                    \exp\left\{-\frac{n\gamma t}{1+2t}\right\}.
\end{align*}
Substituting these into~\eqref{eq:laplace} and changing variables to
$u=2t/(1+2t)$ gives the slightly tidier expression
\bqq
J_{n,\rho}=\frac{\Gamma((1+\rho)d)}{\Gamma(d)\Gamma(\rho d)}I,
\label{eq:integral}
\eqq
where
\bqq
\begin{split}
 I&=\int_0^1 u^{\rho d-1}(1-u)^{d-1/2}e^{-(d+1/2)\gamma u}\,du\\
  &=\int_0^1 e^{dG_{\rho,\gamma}(u)}
                   \frac{e^{-\gamma u/2}}{u\sqrt{1-u}}\,du.
\end{split}
\label{eq:I}
\eqq

Writing $\ell(v):=\log_e v-v+1$ for $v>0$, simple algebra 
using~\eqref{eq:beta} gives the identity
$$
 G_{\rho,\gamma}(u)-G_{\rho,\gamma}(\beta_\rho)
 =\rho\ell\left(\frac {u}{\beta_\rho}\right)
       +\ell\left(\frac{1-u}{1-\beta_\rho}\right).
$$
By straightforward calculus, for $d>1/2$,
\begin{equation*}
 \sup_{w>0}w^{-1/2}e^{d\ell(w)}
 =e^{1/2}\left(1-\frac{1}{2d}\right)^{d-1/2}\leq e^{1/2}.
\end{equation*}
For $d=1/2$, the expression inside the supremum is $e^{(1-w)/2}$,
so the same upper bound holds. Applying it 
in~\eqref{eq:I} 
with
$w=(1-u)/(1-\beta_\rho)$,
and using the obvious inequality
$e^{-\gamma u/2}\leq 1$, we obtain
\begin{equation*}
 I\leq\frac{e^{dG_{\rho,\gamma}(\beta_\rho)+1/2}}{\sqrt{1-\beta_\rho}}
       \int_0^1\exp\left\{\alpha\ell\left(\frac {u}{\beta_\rho}\right)
                  \right\}\frac{du}{u}.
\end{equation*}
Substituting $v=u/\beta_\rho$ and integrating over all of $(0,\infty)$,
the last integral is bounded above by
\begin{equation*}
 \int_0^\infty e^{\alpha\ell(v)}\frac{dv}{v}
 =e^\alpha\int_0^\infty v^{\alpha-1}e^{-\alpha v}\,dv
 =\frac{e^\alpha\Gamma(\alpha)}{\alpha^\alpha}.
\end{equation*}
Noting that the two $\Gamma(\alpha)$ terms cancel
in~\eqref{eq:integral}, we get
\bqq
\label{eq:Jintermediate}
J_{n,\rho}\leq
\frac{e^{dG_{\rho,\gamma}(\beta_\rho)+1/2}}{\sqrt{1-\beta_\rho}}
\frac{\Gamma((1+\rho)d)e^{\rho d}}{\Gamma(d)(\rho d)^{\rho d}}.
\eqq

Recall 
the standard real-argument Stirling-Robbins
bounds~\cite[Section~5.11~(ii)]{NIST:DLMF2}, 
$$
 \sqrt{2\pi}\,x^{x-1/2}e^{-x}\leq\Gamma(x)
 \leq\sqrt{2\pi}\,x^{x-1/2}e^{-x+1/(12x)},
$$
for all $x>0$.
Using both bounds gives
\begin{equation*}
 \frac{\Gamma((1+\rho)d)e^{\rho d}}{\Gamma(d)(\rho d)^{\rho d}}
 \leq\frac{1}{\sqrt{1+\rho}}
        \exp\left\{dg(\rho)+\frac{1}{12(1+\rho)d}\right\},
\end{equation*}
and combining this with~\eqref{eq:Jintermediate} and~\eqref{eq:E0alt}
yields
$$
J_{n,\rho}\leq
\frac{\exp\{\frac12+\frac1{12(1+\rho)d}\}}
      {\sqrt{(1+\rho)(1-\beta_\rho)}}
e^{-(n-1)E_0(\rho,\gamma)}.
$$
Finally, noting that by~\eqref{eq:betabounds} the denominator is at 
least~1, and that for
$d\geq 1/2$ the remaining constant is at most $e^{2/3}<2$,
completes the proof of the lemma.
\qed

\medskip

We are now ready to complete the proof of the theorem.
First
substituting the result of Lemma~\ref{lem:moment}
into~\eqref{eq:Jboundstart} we obtain
\bqq
 \BBE[P_e(C_n)]\leq
 [1+2^{1-\rho}e^{E_0(\rho,\gamma)}]
 e^{-n[E_0(\rho,\gamma)-\rho R]}, \label{eq:probAWGN1rho}
\eqq
and then taking $\rho=\rho^*>0$ and choosing a codebook 
whose error probability does not exceed the ensemble mean 
proves~\eqref{eq:probAWGN1} for $n\geq 2$.
For $n=1$, its right-hand side is at least
$2^{1-\rho^*}e^{\rho^*R}>1$, so the bound is immediate.

Finally,~\eqref{eq:E0bounds} gives
$$D(R,\gamma)\leq 1+2e^{C(\gamma)}=1+2\sqrt{1+\gamma},$$
proving the theorem.
\qed

\subsection{Proof of Theorem~\ref{thm:probAWGN2}}
\label{app:probAWGN2}

We treat the two regimes $\gamma<\gamma_0(R)$ and
$\gamma\geq \gamma_0(R)$ separately. 
When $\gamma<\gamma_0(R)$,~\eqref{eq:probAWGN2} follows 
immediately from~\eqref{eq:probAWGN1snrconstant} in 
Theorem~\ref{thm:probAWGN1}.

Now suppose $\gamma\geq \gamma_0(R)$. We use a standard volumetric
packing argument, which can be seen as a Euclidean analogue of Gilbert's
greedy construction~\cite[Theorem~1]{gilbert:52}; see 
also~\cite[Lemma~4.2.6 and Proposition~4.2.12]{vershynin:book}. Combining this
construction with the Gaussian pairwise-error union bound, as 
in~\cite[Theorem~3]{gilbert:52}, will provide
the claimed high-SNR estimate.

Fix $n\geq 1$ and write $r:=\sqrt{nP}$ and $\delta:=re^{-R}$.
Choose a finite maximal set of points $\widehat C_n$ in the closed
Euclidean ball of radius $r$ in $\RL^n$, with all pairwise distances at
least $\delta$ (such a set always exists by compactness). By maximality, the
balls of radius $\delta$ centered at its points cover the ball of radius
$r$, so comparing volumes gives
$|\widehat C_n|\delta^n\geq r^n$, i.e.,
$$|\widehat C_n|\geq (r/\delta)^n=e^{nR}.$$
Select a subset $ C_n$ 
of $\widehat C_n$ of size $M=\lceil e^{nR}\rceil$. Each of its
codewords satisfies the individual power constraint, and its minimum
distance is at least $\delta$.

Under maximum likelihood decoding, the pairwise error probability between
distinct codewords, $x^n(i)$ and $x^n(j)$, say, is
\begin{equation*}
 Q\left(\frac{\|x^n(i)-x^n(j)\|}{2\sigma}\right)
 \leq\exp\left\{-\frac{\|x^n(i)-x^n(j)\|^2}{8\sigma^2}\right\},
\end{equation*}
where, as before, $Q$ is the standard Gaussian tail function,
and $\|\cdot\|$ denotes the Euclidean norm on $\RL^n$.
Since $M-1<e^{nR}$, the union bound gives
\bqq
\label{eq:packing}
 \epsilon^*(n,R,\sigma^2,P)
 \leq\exp\left\{-n\left[\frac{\gamma e^{-2R}}{8}-R\right]\right\}.
\eqq
Note that the high-SNR assumption has not been used yet,
so this bound holds for every
$R>0$, $\gamma>0$ and $n\geq 1$.

Now we only need to check that 
\bqq
C(\gamma)\leq\gamma e^{-2R}/8,
\label{eq:last}
\eqq
for $\gamma\geq \gamma_0(R)$. Let
$F(\gamma):=\gamma e^{-2R}/8-\frac{1}{2}\log_e(1+\gamma)$.
For $\gamma\geq \gamma_0(R)$ we have $1+\gamma\geq 4e^{2R}$, hence
\begin{equation*}
 F'(\gamma)=\frac{e^{-2R}}{8}-\frac{1}{2(1+\gamma)}\geq 0,
\end{equation*}
and so it suffices to check~\eqref{eq:last}
at $\gamma_0(R)$. To that end, note that we have
$1+\gamma_0(R)\leq 17(1+R)e^{2R}$, and therefore,
\begin{align*}
 C(\gamma_0(R))
 &\leq\tfrac12\log_e 17+\tfrac12\log_e(1+R)+R\\
 &<\tfrac32+\tfrac32R\leq 2(1+R)\\
 &=\frac{\gamma_0(R)e^{-2R}}{8}.
\end{align*}
This establishes~\eqref{eq:last} and combining it 
with~\eqref{eq:packing} and~\eqref{eq:Erbounds} gives
\begin{equation*}
 \epsilon^*(n,R,\sigma^2,P)
 \leq e^{-n[C(\gamma)-R]}\leq e^{-n E_r(R,\gamma)},
\end{equation*}
which completes the proof of the theorem.

\subsection{Proof of Corollary~\ref{cor:probAWGN3}} 
\label{app:probAWGN3}

The first bound, in~\eqref{eq:probAWGN3.1},
immediately follows from the bound derived in~\eqref{eq:probAWGN1rho} 
in the proof or Theorem~\ref{thm:probAWGN1}, upon taking
$\rho = 1$ and $R = \frac{1}{2}C_{1/2}(\gamma)$,
for $n\geq 2$. The bound is trivial for $n=1$.

The second bound of the corollary,
in~\eqref{eq:probAWGN3.2}, 
will follow 
by an application of the bound~\eqref{eq:probAWGN2bis} in
Theorem~\ref{thm:probAWGN2}, once we show that,
with $R=\frac{1}{2}C_{1/2}(\gamma)$,
taking
$\gamma\geq 1300$ implies that
$\gamma\geq\gamma_0(R)$, i.e., that
\be
\gamma - \gamma_0\Big(\frac{1}{2}C_{1/2}(\gamma)\Big)=
\gamma - 16\Big(1+\frac{1}{2}C_{1/2}(\gamma)\Big)e^{C_{1/2}(\gamma)}
\geq 0.
\label{eq:cor}
\ee
To that end, we recall the expression for
$C_{1/2}(\gamma)$ in Proposition~\ref{prop:AWGNcutoff}, 
and note that the  
straightforward bounds $\gamma \leq \Delta \leq \gamma + \frac{2}{\gamma}$, 
imply, for $\gamma \geq 2$,
\begin{align*}
2 C_{1/2}(\gamma) &= \left( 1 + \frac{\gamma - \Delta}{2} \right) 
+ \log_e \left( \frac{1}{2} + \frac{\Delta}{4} \right) \\
&\leq 1 + \log_e \left( \frac{2+\gamma+\frac{2}{\gamma}}{4} \right) \\
&\leq U(\gamma):= 1 + \log_e \left( \frac{3+\gamma}{4} \right).
\end{align*}
Since $\gamma_0(R)$ is nondecreasing,
this gives,
\begin{align*}
\gamma - \gamma_0\left(\frac{1}{2}C_{1/2}(\gamma)\right) 
&\geq \gamma - \gamma_0\left(\frac{1}{4}U(\gamma)\right) \\
&= m(\gamma):=\gamma - \frac{\sqrt{e}}{2}
\left(20 + 4 \log_e \left( \frac{\gamma+3}{4} \right)\right)\sqrt{\gamma+3}.
\end{align*}
Simple calculus then shows that
$m'(\gamma)\geq 0$ for $\gamma\geq 1300$,
so in that range,
$$\gamma - \gamma_0\left(\frac{1}{2}C_{1/2}(\gamma)\right) 
\geq m(1300)> 16.14>0.$$
This establishes~\eqref{eq:cor} and completes the proof.
\qed

\end{appendices}

%

\bibliographystyle{plain}
\bibliography{ik}

\end{document}